\documentclass[twoside,twocolumn,9pt]{article}
\usepackage{extsizes}
\usepackage[super,sort&compress,comma]{natbib} 
\usepackage[version=3]{mhchem}
\usepackage[left=1.5cm, right=1.5cm, top=1.785cm, bottom=2.0cm]{geometry}
\usepackage{balance}
\usepackage{mathptmx}
\usepackage{sectsty}
\usepackage{graphicx} 
\usepackage{lastpage}
\usepackage[format=plain,justification=justified,singlelinecheck=false,font={stretch=1.125,small,sf},labelfont=bf,labelsep=space]{caption}
\usepackage{float}
\usepackage{fancyhdr}
\usepackage{fnpos}
\usepackage[english]{babel}
\addto{\captionsenglish}{%
  \renewcommand{\refname}{Notes and references}
}
\usepackage{array}
\usepackage{droidsans}
\usepackage{charter}
\usepackage[T1]{fontenc}
\usepackage[usenames,dvipsnames]{xcolor}
\usepackage{setspace}
\usepackage[compact]{titlesec}

\DeclareFontSeriesDefault[sf]{bf}{bx}

\usepackage[hidelinks]{hyperref}

\definecolor{cream}{RGB}{222,217,201}

\usepackage{units,bm,amssymb}

\newcommand{\dpart}[2]{\frac{\partial #1}{\partial #2}}

\begin{document}

\pagestyle{fancy}
\thispagestyle{plain}
\fancypagestyle{plain}{
\renewcommand{\headrulewidth}{0pt}
}

\makeFNbottom
\makeatletter
\renewcommand\LARGE{\@setfontsize\LARGE{15pt}{17}}
\renewcommand\Large{\@setfontsize\Large{12pt}{14}}
\renewcommand\large{\@setfontsize\large{10pt}{12}}
\renewcommand\footnotesize{\@setfontsize\footnotesize{7pt}{10}}
\makeatother

\renewcommand{\thefootnote}{\fnsymbol{footnote}}
\renewcommand\footnoterule{\vspace*{1pt}%
\color{cream}\hrule width 3.5in height 0.4pt \color{black}\vspace*{5pt}} 
\setcounter{secnumdepth}{5}

\makeatletter 
\renewcommand\@biblabel[1]{#1}            
\renewcommand\@makefntext[1]%
{\noindent\makebox[0pt][r]{\@thefnmark\,}#1}
\makeatother 
\renewcommand{\figurename}{\small{Fig.}~}
\sectionfont{\sffamily\Large}
\subsectionfont{\normalsize}
\subsubsectionfont{\bf}
\setstretch{1.125}
\setlength{\skip\footins}{0.8cm}
\setlength{\footnotesep}{0.25cm}
\setlength{\jot}{10pt}
\titlespacing*{\section}{0pt}{4pt}{4pt}
\titlespacing*{\subsection}{0pt}{15pt}{1pt}

\fancyfoot{}
\fancyfoot[LO,RE]{\vspace{-7.1pt}\includegraphics[height=9pt]{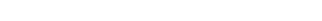}}
\fancyfoot[CO]{\vspace{-7.1pt}\hspace{13.2cm}\includegraphics{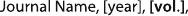}}
\fancyfoot[CE]{\vspace{-7.2pt}\hspace{-14.2cm}\includegraphics{head_foot/RF}}
\fancyfoot[RO]{\footnotesize{\sffamily{1--\pageref{LastPage} ~\textbar  \hspace{2pt}\thepage}}}
\fancyfoot[LE]{\footnotesize{\sffamily{\thepage~\textbar\hspace{3.45cm} 1--\pageref{LastPage}}}}
\fancyhead{}
\renewcommand{\headrulewidth}{0pt} 
\renewcommand{\footrulewidth}{0pt}
\setlength{\arrayrulewidth}{1pt}
\setlength{\columnsep}{6.5mm}
\setlength\bibsep{1pt}

\makeatletter 
\newlength{\figrulesep} 
\setlength{\figrulesep}{0.5\textfloatsep} 

\newcommand{\topfigrule}{\vspace*{-1pt}%
\noindent{\color{cream}\rule[-\figrulesep]{\columnwidth}{1.5pt}} }

\newcommand{\botfigrule}{\vspace*{-2pt}%
\noindent{\color{cream}\rule[\figrulesep]{\columnwidth}{1.5pt}} }

\newcommand{\dblfigrule}{\vspace*{-1pt}%
\noindent{\color{cream}\rule[-\figrulesep]{\textwidth}{1.5pt}} }

\makeatother

\twocolumn[
  \begin{@twocolumnfalse}
{\includegraphics[height=30pt]{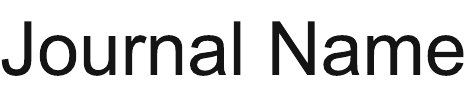}\hfill\raisebox{0pt}[0pt][0pt]{\includegraphics[height=55pt]{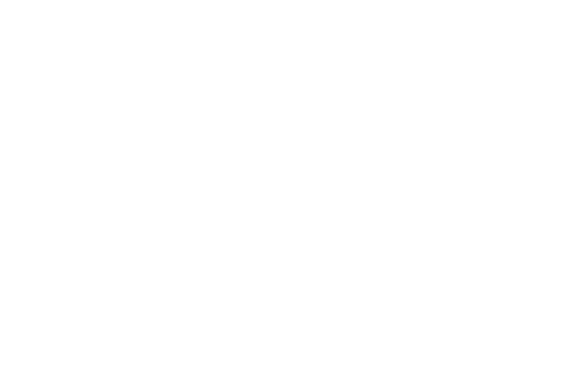}}\\[1ex]
\includegraphics[width=18.5cm]{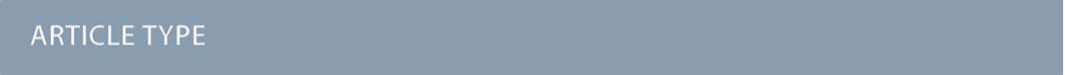}}\par
\vspace{1em}
\sffamily
\begin{tabular}{m{4.5cm} p{13.5cm} }

\includegraphics{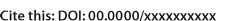} & \noindent\LARGE{\textbf{Mean-field model for the bubble size distribution in coarsening wet foams}} \\
\vspace{0.3cm} & \vspace{0.3cm} \\

 & \noindent\large{Jacob Morgan$^{\ast}$\textit{$^{a}$} and Simon Cox\textit{$^{a}$}} \\

\includegraphics{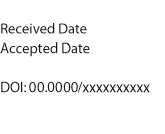} & \noindent\normalsize{Aqueous foams are subject to coarsening, whereby gas from the bubbles diffuses through the liquid phase. Gas is preferentially transported from small to large bubbles, resulting in a gradual decrease of the number of bubbles and an increase in the average bubble size. Coarsening foams are expected to approach a scaling state at late times in which their statistical properties are invariant. However, a model predicting the experimentally observed bubble-size distribution in the scaling state of foams with moderate liquid content, as a function of the liquid fraction $\phi$, has not yet been developed. To this end, we propose a three-dimensional mean-field bubble growth law for foams without inter-bubble adhesion, validated against bubble-scale simulations, and use it to derive a prediction of the scaling-state bubble-size distribution for any $\phi$ from zero up to the unjamming transition $\phi_\text{c} \approx 36\%$. We verify that the derived scaling state is approached from a variety of initial conditions using mean-field simulations implementing the proposed growth law. Comparing our predicted bubble-size distribution with previous simulations and experimental results, we likewise find a large population of small bubbles when $\phi > 0$, but there are qualitative differences from prior results which we attribute to the absence of rattlers, i.e.\ bubbles not pressed into contact with their neighbours, in our model.} \\

\end{tabular}

 \end{@twocolumnfalse} \vspace{0.6cm}

  ]

\renewcommand*\rmdefault{bch}\normalfont\upshape
\rmfamily
\section*{}
\vspace{-1cm}


\footnotetext{\textit{$^{a}$~Department of Mathematics, Aberystwyth University, Aberystwyth, UK.}}
\footnotetext{\textit{$^{\ast}$~E-mail: jam164@aber.ac.uk}}



\section{Introduction}\label{sec:intro}
	
	Aqueous foams --- packings of gas bubbles in liquid\cite{1999-weaire,2013-cantat} --- are of theoretical importance as complex fluids whose properties arise from their well-characterised small-scale structure.\cite{2023-stewart} Foams also have many industrial applications, including fire suppression, ore separation, foods, and drinks.\cite{2012-martin,2012-stevenson,1999-weaire} However, a foam's lifetime is limited by film rupture, gravitational drainage, and \emph{coarsening}.\cite{2013-cantat} Our focus is on the last of these instabilities, which occurs due to solubility of the gas phase in the liquid: dissolved gas diffuses between bubbles, preferentially transferring from high-pressure bubbles to those at lower pressure.\cite{1978-lemlich,2017-schimming} Small bubbles tend to have higher pressure (the pressure of isolated spherical bubbles is inversely proportional to their radius by the Young-Laplace law\cite{1999-weaire}), and thus shrink until they disappear, while large bubbles grow and the mean bubble size increases.\cite{1978-lemlich,2013-cantat} We neglect film rupture and drainage henceforth, noting that they can be mitigated in experiments, albeit with difficulty in the latter case.\cite{2013-isert,2013-roth,2021-born}
	
	The effects of coarsening are fairly well characterised in the dry and wet limits of a foam's liquid content,\cite{1959-lifshitz,1961-wagner,1986-mullins,1992-glazier,1993-stavans,2006-thomas,2010-lambert,2013-cantat} which is measured by the \emph{liquid fraction} $\phi$; i.e.\ the ratio of liquid volume to the foam's total volume. The foam approaches a \emph{scaling state} in which its statistical properties do not change with time, except for a scaling arising from growth of the average bubble size.\cite{1959-lifshitz,1986-mullins,2010-lambert,2013-cantat} Let $R$ denote the \emph{effective radius} of a bubble; i.e.\ the radius of a sphere with the same volume (or that of a circle of the same area in 2D).\cite{1999-weaire} Then the growth exponent $\alpha$ with which the mean radius $\langle R \rangle$ (along with other averages\cite{2023-pasquet-b} of $R$) increases with time $t$ in the scaling state, via $\langle R \rangle \sim t^\alpha$, is $1/2$ in the dry limit $\phi \to 0$ and $1/3$ in the wet limit $\phi \to 1$.\cite{1959-lifshitz,2010-lambert,2013-cantat} However, real foams lie between these limits, and moderately-wet foams arise in applications such as fire suppression and ore separation.\cite{2012-martin,2012-stevenson}
	
	Coarsening at moderate $\phi$ is not yet well characterised. Experiments and simulations indicate that the foam still approaches a scaling state,\cite{2012-fortuna,2015-thomas,2018-khakalo,2023-galvani} although they are not all consistent regarding the form of the crossover in $\alpha$ from $1/2$ to $1/3$ as $\phi$ increases,\cite{2012-fortuna,2013-isert,2015-thomas,2018-khakalo,2023-pasquet-b} and unexpected bubble-size distributions have been observed in simulations\cite{2018-khakalo} and in recent experiments on the International Space Station (ISS).\cite{2023-galvani} The observed distributions exhibit a large population of small bubbles at moderate $\phi$, which has been interpreted as resulting from the small shrinkage rate of \emph{rattlers} (also termed \emph{roamers}); i.e.\ small bubbles, which have been observed directly, that are either out of contact with their neighbours, or that have contacts due only to inter-bubble adhesion.\cite{2018-khakalo,2023-galvani,2025-galvani} However, rigorous theoretical predictions of the variation of $\alpha$ with $\phi$, and of the scaling-state bubble-size distribution, have yet to be developed.
	
	Progress on predicting $\alpha$ has recently been made by \citeauthor{2023-durian-pre-n-yo},\cite{2023-durian-pre-n-yo} who derived an approximate growth law for the average bubble radius, with contributions from gas transfer through thin  films separating the bubbles, and through the bulk liquid in the foam. The growth exponent $\alpha$ was taken to be a weighted average of these contributions, with the typical film area obtained by fitting to experimental data.\cite{2023-durian-pre-n-yo} Encouraging agreement with experiments was found,\cite{2013-isert,2023-pasquet-b,2023-durian-pre-n-yo} but there is considerable scope for improving the argument's rigour, and possibly for avoiding the use of free parameters.
	
	Our focus in the present work is instead to derive an approximation of the scaling-state distribution as a function of $\phi$. We apply standard techniques\cite{1959-lifshitz,1983-marqusee} to derive the distribution predicted by the three-dimensional (3D) version of a mean-field bubble growth law we proposed in prior work.\cite{2024-morgan} This is a \emph{border-blocking} growth law; i.e.\ it accounts only for gas transfer through a bubble's contact films, omitting that through the bulk liquid,\cite{1991-bolton,2013-roth} and hence corresponds to the limit in which the ratio of film thickness $h$ to bubble size tends to zero.\cite{2017-schimming,2023-durian-pre-n-yo,2023-pasquet-b} This would appear to be a natural first case to consider: as a foam coarsens, $\langle R \rangle$ grows while $h$ is usually expected to remain approximately constant,\cite{1999-weaire,2013-roth,2023-pasquet-b} and so the limit $h / \langle R \rangle \to 0$ should be approached at late times. A border-blocking model might therefore be expected to describe the eventual coarsening dynamics of a real foam, and thus its scaling state (analogous behaviour is understood to occur in alloy coarsening, where the dynamics is eventually dominated by bulk diffusion instead\cite{1979-white,2022-li}). This argument assumes that there are films in the foam,\footnote[2]{There may be other exceptions. For example, it is conceivable that a population of non-evolving rattlers in a border-blocking model\cite{1999-gardiner} could preclude a scaling state. This possibility is discussed further in Section~\ref{sec:discussion}.} and thus that the foam is either jammed, or flocculated due to bubble adhesion.\cite{1968-princen,2018-cox,2023-pasquet-b} For simplicity, and as the natural starting point for developing theory, we consider only foams with zero adhesion in the present work; i.e.\ with a contact angle $\theta = 0$ between film and bulk-liquid interfaces.\cite{1979-princen} Therefore, we restrict our attention to jammed foams, with $\phi < \phi_\text{c}$, where $\phi_\text{c} \approx 36\%$ is the unjamming transition in 3D.\cite{2013-cantat}

	Mean-field models are widely used to provide tractable approximations of coarsening systems.\cite{1978-lemlich,1997-yarranton,2023-chieco,2023-pasquet-b} In the dry limit $\phi = 0$, \emph{Lemlich's law}\cite{1961-wagner,1965-hillert,1978-lemlich} approximates the growth rate of a bubble as follows. Let the bubble have effective radius $R$, and let $\gamma$ be the liquid/gas surface tension, $D$ a gas diffusion coefficient, $H$ Henry's constant,\cite{2013-cantat} and $h$ the film thickness. Let $R_\text{c}$ denote the \emph{critical bubble radius}, which is the radius of bubbles which neither grow nor shrink under coarsening,\cite{1959-lifshitz,1978-lemlich} and which is set by the condition that the total gas volume is conserved; i.e.\ that the mean volumetric bubble growth rate satisfies\cite{1978-lemlich,2013-cantat} $\langle \dot{V} \rangle \equiv \langle d V / d t \rangle = 0$. Lemlich's law is then\cite{1961-wagner,1965-hillert,1978-lemlich}
	\begin{align}
		\frac{d R}{d t} = \frac{2 \gamma D H}{h} \left(\frac{1}{R_\text{c}} - 	
			\frac{1}{R}\right) , \label{eqn:growth-law-lemlich}
	\end{align}
	where\cite{1978-lemlich,2013-cantat} $R_\text{c} = \langle R^2 \rangle / \langle R \rangle \equiv R_{2 1}$. This law is approximate due to mean-field assumptions made during its derivation. In the wet limit $\phi \to 1$, for which all gas flow is through the bulk liquid (\emph{Ostwald ripening}),\cite{2013-cantat} the \emph{Lifshitz-Slyozov-Wagner (LSW) law} gives\cite{1959-lifshitz,1961-wagner}
	\begin{align}
		\frac{d R}{d t} = \frac{2 \gamma D H}{R} \left(\frac{1}{R_\text{c}} - 	
		\frac{1}{R}\right) , \label{eqn:growth-law-lsw}
	\end{align}
	where imposing $\langle \dot{V} \rangle = 4 \pi \langle R^2 \dot{R} \rangle = 0$ gives\cite{1961-wagner} $R_\text{c} = \langle R \rangle$.
	
	Much prior work\cite{1972-ardell,1997-yarranton,2013-streitenberger,2022-li} has been done to develop mean-field growth laws for intermediate values of $\phi$. However, to our knowledge, most of these studies are focussed on more general coarsening systems, rather than foams specifically, and thus do not typically model the films between contacting bubbles. Gas transfer in foams is most efficient through films due to their small thickness,\cite{2013-cantat} and so it seems necessary to approximate film sizes to obtain an effective foam coarsening model. Progress has been made by \citeauthor{2023-pasquet-b},\cite{2023-pasquet-b} who obtain a correction to the prefactor in eqn~\eqref{eqn:growth-law-lemlich} for $\phi > 0$ using an approximation for film area as a function of $\phi$ derived by \citeauthor{2021-hohler},\cite{2021-hohler} as we shall discuss in Section~\ref{sec:growth-law-bb}. While their model predicts the variation of the coarsening rate with $\phi$, it is not able to predict the effect of $\phi$ on the bubble size distribution, since their growth law is a constant multiple of Lemlich's law \eqref{eqn:growth-law-lemlich} at fixed $\phi$ (see Section~\ref{sec:growth-law-bb}), thus justifying development of the more detailed model that we describe below.
	
	In Section~\ref{sec:growth-law-bb}, we define our proposed growth law, and compare its predictions with our previous bubble-scale simulations.\cite{2024-morgan,2025-morgan-thesis} Then, in Section~\ref{sub:reb:analysis}, we adapt standard techniques\cite{1959-lifshitz,1983-marqusee} to derive the scaling-state bubble-size distribution predicted by this growth law. Next, in Section~\ref{sec:sims}, we give the results of mean-field simulations using our proposed growth law to check that the derived scaling state is approached from various initial conditions. Our simulation methods are adapted from those of \citet{1997-de-smet} We then compare the derived distribution with previous experiments\cite{2023-galvani,2025-galvani} and simulations\cite{2018-khakalo} in Section~\ref{sec:discussion}. We conclude in Section~\ref{sec:conclusion}, and discuss simulation convergence and apparent exceptions to the universality of our derived scaling state (comparable to apparent exceptions previously observed for other mean-field laws\cite{1989-brown,2022-li,2022-svoboda}) in the \hyperref[sec:app]{Appendix}.
	
\section{Mean-field growth law for wet foams}\label{sec:growth-law-bb}
	
	The border-blocking growth rate of a bubble in a 3D wet foam is known exactly under standard assumptions (including that the thin film thickness $h$ is the same for all contact films), although we will make mean-field approximations below in order to develop a tractable model. Let the considered bubble have effective radius $R$ (i.e.\ the radius of a sphere with equal volume\cite{1999-weaire}), pressure $p$, and $n$ contacting neighbours (and hence $n$ films). Let the bubble's $k^\text{th}$ contact film have area $A_k$ and adjoin a bubble of pressure $p_k$. Then the bubble's border-blocking growth rate is exactly\cite{2013-cantat}
	\begin{align}
		\frac{d R}{d t} = \frac{D H}{h} \, \frac{1}{4 \pi R^2} \, \sum_{k=1}^n (p_k - p) \, A_k . 
			\label{eqn:growth-law-bb-exact}
	\end{align}
	In 2D, the equivalent growth law is\cite{1991-bolton}
	\begin{align}
		\frac{d R}{d t} = \frac{D H}{h} \, \frac{1}{2 \pi R} \, \sum_{k=1}^n (p_k - p) \, L_k \qquad \text{(in 2D)} ,
		\label{eqn:growth-law-bb-2d-exact}
	\end{align}
	where $L_k$ is the length of the film between the bubble and its $k^\text{th}$ neighbour. Due to the dependence of eqn~\eqref{eqn:growth-law-bb-exact} on multiple bubble properties (not just $R$, for example), its implications for the evolution of a coarsening foam as a whole are not clear, inviting further approximations. \citet{2023-pasquet-b} proposed a mean-field border-blocking growth law which generalises eqn~\eqref{eqn:growth-law-lemlich}\cite{1978-lemlich} to wet foams. It is based on the following approximation, which was derived by \citeauthor{2021-hohler},\cite{2021-hohler} for the ratio of total contact film area $A$ to surface area $S$ of bubbles in monodisperse 3D foams. Let the foam's osmotic pressure\cite{1979-princen,1986-princen} be $\Pi_\text{O}$, and define the scaled osmotic pressure $\hat{\Pi} = \Pi_\text{O} (R_{3 2} / \gamma) / (1 - \phi)$ for brevity (adapted from \citet{1986-princen}), where\cite{2013-cantat} $R_{3 2} \equiv \langle R^3 \rangle / \langle R^2 \rangle$. Then\cite{2021-hohler}
	\begin{align}
		\frac{A}{S} \approx \frac{\hat{\Pi}}{2 + \hat{\Pi}} . \label{eqn:film-area-hohler}
	\end{align}
	Following \citeauthor{2023-pasquet-b},\cite{2023-pasquet-b} let us adapt the derivation\cite{1978-lemlich} of Lemlich's law~\eqref{eqn:growth-law-lemlich}. We simplify eqn~\eqref{eqn:growth-law-bb-exact} by making the mean-field approximation that all neighbours and films of the considered bubble are identical, and take the bubble to have pressure $p = 2 \gamma / R$; i.e.\ that of an isolated spherical bubble,\cite{1978-lemlich} by the Young-Laplace law.\cite{1999-weaire} The equal neighbour pressures are written $p_k = 2 \gamma / R_\text{c}$ (anticipating that $R_\text{c}$ will serve as the critical radius), and the bubble's total film area is approximated using eqn~\eqref{eqn:film-area-hohler} (taking\cite{2004-kraynik} $S \approx 4\pi R^2$). \citet{2023-pasquet-b} thereby obtained an approximate growth law of the form
	\begin{align}
		\frac{d R}{d t} = \frac{2 \gamma D H}{h} \left(\frac{1}{R_\text{c}} - \frac{1}{R}\right)
			\frac{\hat{\Pi}}{2 + \hat{\Pi}} , \label{eqn:growth-law-pasquet}
	\end{align}
	where $R_\text{c} = R_{2 1}$ since the law has the same functional form as eqn~\eqref{eqn:growth-law-lemlich}. While this growth law predicts the rate at which the foam coarsens\cite{2023-pasquet-b} (see Section~\ref{sub:reb:sims} below), it is not suitable for approximating the variation of the scaling-state bubble-size distribution with $\phi$ --- the right-most factor, approximating the bubble's film coverage, scales all bubble growth rates equally, and so the same scaling-state distribution, that of Lemlich's law~\eqref{eqn:growth-law-lemlich},\cite{1961-wagner,1965-hillert,1985-markworth} is obtained for all $\phi < \phi_\text{c}$ (beyond which $\hat{\Pi} = 0$ and there is no coarsening in the border-blocking model\cite{2017-schimming,2021-hohler}). This suggests the development of growth laws which account for the correlation between bubble pressure and film area, an approach which was proposed by \citet{2023-pasquet-b}
	
	In ref.~\citenum{2024-morgan}, we followed this approach for two-dimensional (2D) wet foams by developing mean-field approximations for bubble pressure $p$ and the ratio of a bubble's total film length $L_\text{f}$ to its perimeter $P$, which depend only on bubble radius $R$ (in the case of the non-adhesive foams we consider here). Thus, the correlation between $p$ and $L_\text{f}/P$ is approximated through their mutual dependence on $R$. These approximations for $p$ and $L_\text{f}/P$ are generalisations of two results from \citet{2021-hohler} --- a relation between the foam's capillary and osmotic pressures, and eqn~\eqref{eqn:film-area-hohler} --- to individual bubbles in polydisperse foams, and are derived using similar techniques (but are subject to additional simplifying assumptions). In ref.~\citenum{2024-morgan}, we used the pressure and film length approximations, in a similar manner to the derivation\cite{1978-lemlich} of eqns~\eqref{eqn:growth-law-lemlich} and~\eqref{eqn:growth-law-pasquet}, to derive the following approximate mean-field border-blocking growth law, which incorporates the variation of film coverage with bubble size. Let $\bar{\Pi} = \Pi_\text{O} (R_{2 1} / \gamma) / (1 - \phi)$ be a 2D equivalent of $\hat{\Pi}$, and let $\bar{R} = R / R_{2 1}$. Then\cite{2024-morgan}
	\begin{align}
		\frac{d R}{d t} = \frac{\gamma D H}{h} \left(\frac{1}{R_\text{c}} - \frac{1}{R}\right)
		\frac{2 \bar{\Pi} \bar{R}}{1 + (1 + 2 \bar{\Pi}) \bar{R}} \qquad \text{(in 2D)} , \label{eqn:growth-law-bb-2d}
	\end{align}
	where $R_\text{c}$ is the critical radius. A 3D equivalent of this result can be derived using very similar arguments, which are given in ref.~\citenum{2025-morgan-thesis}. The resulting 3D mean-field approximation for an individual bubble's pressure $p$ is
	\begin{align}
		p \approx \frac{\Pi_\text{O}}{1 - \phi} + \frac{2 \gamma}{R} . \label{eqn:press-approx}
	\end{align}
	The corresponding approximation for the ratio of a bubble's total film area $A$ to its surface area $S$ (i.e.\ the fraction of its surface in contact with other bubbles) is
	\begin{align}
		\frac{A}{S} \approx \frac{\hat{\Pi} \hat{R}}{1 + (1 + \hat{\Pi}) \hat{R}} ,
			\label{eqn:film-area-approx}
	\end{align}
	where $\hat{R} = R / R_{3 2}$. The scaling of radius by $R_{3 2}$ arises from an approximation that all neighbouring bubbles have pressure equal to the capillary pressure $\Pi_\text{c}$ of the foam.\cite{2021-hohler} Other pressures could have been chosen, but we shall see below that the resulting growth law is insensitive to a replacement of $R_{3 2}$ by $R_\text{c}$, which is another average radius. Eqn~\eqref{eqn:film-area-approx} becomes eqn~\eqref{eqn:film-area-hohler}\cite{2021-hohler} in a monodisperse foam (where $\hat{R} = 1$ for all bubbles).
	
	An approximate 3D growth law is then obtained, like eqns~\eqref{eqn:growth-law-lemlich} and~\eqref{eqn:growth-law-pasquet}, by assuming identical films and neighbours in eqn~\eqref{eqn:growth-law-bb-exact}. Eqn~\eqref{eqn:press-approx} is used for the bubble's pressure (instead of the Young-Laplace law for spherical bubbles), and the equal pressures of its neighbours (taken to have radius $R_\text{c}$), while eqn~\eqref{eqn:film-area-approx} gives the total film area $A$ (approximating\cite{2004-kraynik} $S \approx 4 \pi R^2$). Hence, we obtain the approximate 3D mean-field border-blocking growth law
	\begin{align}
		\frac{d R}{d t} = \frac{2 \gamma D H}{h} \left(\frac{1}{R_\text{c}} - \frac{1}{R}\right)
			\frac{\hat{\Pi} \hat{R}}{1 + (1 + \hat{\Pi}) \hat{R}} , \label{eqn:growth-law-bb}
	\end{align}
	where $R_\text{c}$ is the critical radius. Comparing with eqn~\eqref{eqn:growth-law-pasquet}, we see that there is an additional dependence on bubble size $R$, which is intended to approximate the variation of film size with $R$, and thus more accurately describe a bubble's growth rate (still within a mean-field model). As noted above, this additional complexity is required to predict the variation of scaling-state bubble-size distribution with $\phi$ (upon which it is known to depend\cite{2017-khakalo-pre-n,2023-galvani}).
	
	In ref.~\citenum{2024-morgan}, we compared eqn~\eqref{eqn:growth-law-bb-2d} with the growth rates predicted by bubble-scale finite-element simulations of 2D foams. These simulations were implemented in the Surface Evolver,\cite{1992-brakke,2013-brakke} and adapted the approaches of \citet{2014-kahara} and \citeauthor{2018-boromand}:\cite{2018-boromand} every liquid/gas interface in the foam was resolved (requiring a film thickness considerably larger than expected in real static foams), and made to interact with other interfaces through a disjoining pressure. We found that eqn~\eqref{eqn:growth-law-bb-2d} appears to approximate the trend in the simulated border-blocking growth rates (when corrected for variations in simulated bubble film thickness $h$; see below), albeit with a large degree of scatter among individual bubbles.
	
\begin{figure}[h]
		\centering
		\includegraphics[width=\columnwidth]{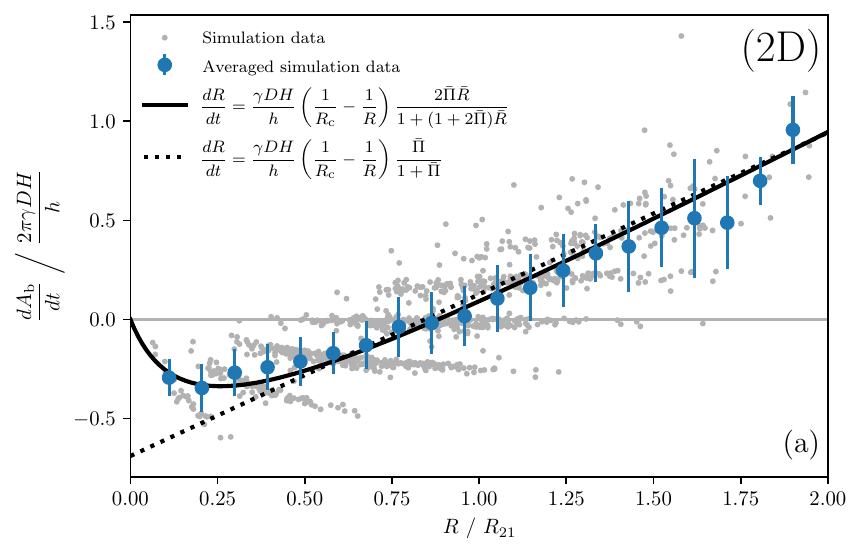}
		\includegraphics[width=\columnwidth]{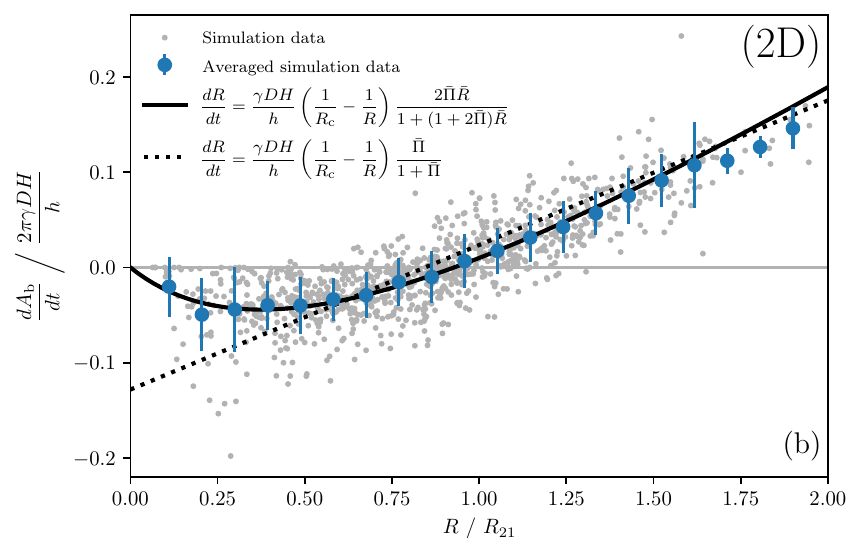}
		\caption{Border-blocking growth rate (expressed as the rate of change in bubble area $A_\text{b} = \pi R^2$) versus effective radius for $1024$ bubbles in 2D simulated foams at (a) $\phi = 2\%$ and (b) $\phi = 10\%$ without inter-bubble adhesion, the simulation data having been taken from our previous study.\cite{2024-morgan,2023-morgan-data} The bubble-area distribution is a compressed exponential fitted to experimental data by \citet{2013-roth} The individual bubble data is shown alongside its binned mean ($20$ equal bins are used; the error bars give the standard deviation within each bin). Comparison is made with eqn~\eqref{eqn:growth-law-bb-2d} and the 2D version of eqn~\eqref{eqn:growth-law-pasquet}, in which $R_\text{c}$ is obtained by numerically solving $\langle d A_\text{b} / d t \rangle = 0$ (i.e.\ conservation of total gas area\cite{1978-lemlich}) and $\bar{\Pi}$ is measured in the simulations.\cite{2024-morgan} The calculation of $d A_\text{b} / d t$ in the simulations is explained in the text.}
		\label{fig:growth-law-2d}
	\end{figure}

	Using the data\cite{2023-morgan-data} from ref.~\citenum{2024-morgan}, we see in Fig.~\ref{fig:growth-law-2d} that eqn~\eqref{eqn:growth-law-bb-2d} is actually in good agreement with the binned mean growth rate. Here, we have used eqn~\eqref{eqn:growth-law-bb-2d-exact} to calculate the border-blocking growth rates of the simulated bubbles (expressed in terms of their effective neighbour number\cite{2012-fortuna,2024-morgan}). Eqn~\eqref{eqn:growth-law-bb-exact} assumes that all films have equal thickness $h$, which is a standard approximation for experimental foams,\cite{1999-weaire,2013-roth,2023-pasquet-b} whereas film thickness varies considerably between bubbles in our simulations for numerical reasons, an effect which would tend to enhance the shrinkage rates of small bubbles (see ref.~\citenum{2024-morgan} for details). However, we do not expect the thickness variations to affect the bubble pressures and film sizes substantially, and so our use of eqn~\eqref{eqn:growth-law-bb-2d-exact} should correct for these variations.
	
	\begin{figure}[h]
		\centering
		\includegraphics[width=\columnwidth]{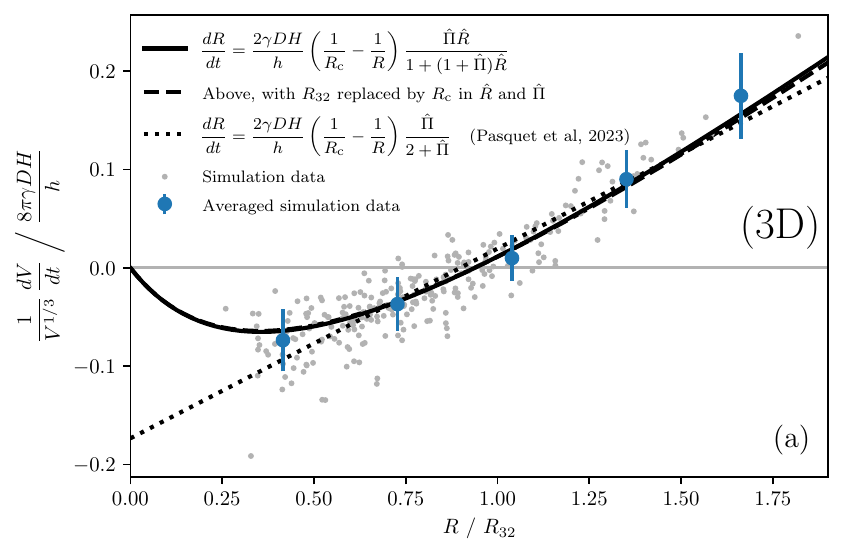}
		\includegraphics[width=\columnwidth]{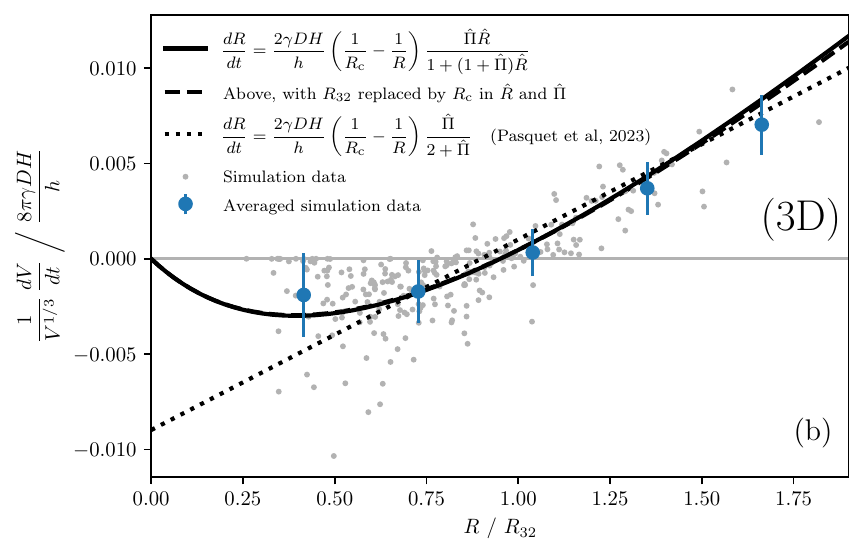}
		\caption{Border-blocking relative growth rate\cite{2013-cantat} (where $V = 4 \pi R^3 / 3$ is the bubble volume) versus effective radius for $256$ bubbles (aggregated from four $64$-bubble simulations at equal $\phi$) in 3D foams, using finite-element simulations described in ref.~\citenum{2025-morgan-thesis}. There is no inter-bubble adhesion, and the liquid fraction is (a) $\phi = 10\%$ and (b) $\phi = 30\%$. The bubble-size distribution\cite{2010-lambert} is lognormal with standard deviation $0.4$ with respect to $R / \langle R \rangle$. The simulation data is plotted as in Fig.~\ref{fig:growth-law-2d}, except $5$ equal bins are used due to the smaller number of bubbles, and is compared with eqns~\eqref{eqn:growth-law-bb} and~\eqref{eqn:growth-law-pasquet};\cite{2023-pasquet-b} where $\hat{\Pi}$ and $R_{3 2}$ are the mean values measured in the aggregated simulations, and $R_\text{c}$ is obtained\cite{1978-lemlich} by numerically solving $\langle d V / d t \rangle = 0$. The means of calculating $d V / d t$ is stated in the text. Comparison is also made to eqn~\eqref{eqn:growth-law-bb} with $R_{3 2}$ replaced by $R_\text{c}$ as described in the text, where $R_\text{c}$ is again calculated by solving $\langle d V / d t \rangle = 0$.}
		\label{fig:growth-law-3d}
	\end{figure} 
	
	In Fig.~\ref{fig:growth-law-2d}, we also see improved agreement over the 2D equivalent\cite{1965-hillert,2024-morgan} of eqn~\eqref{eqn:growth-law-pasquet} for small bubbles, further justifying the more complicated growth law~\eqref{eqn:growth-law-bb-2d}.
	
	We provide a similar test of eqn~\eqref{eqn:growth-law-bb} in Fig.~\ref{fig:growth-law-3d}, using 3D versions of the above finite-element simulations, which are described in detail in ref.~\citenum{2025-morgan-thesis}. These simulations use the same approach as in 2D, that of \citet{2014-kahara} and \citeauthor{2018-boromand},\cite{2018-boromand} and are similar to prior simulations of 3D frictional particles\cite{2021-wang} and biological cells.\cite{2020-van-liedekerke} The simulations are again performed with the Surface Evolver,\cite{2013-brakke} and the initial dry-foam geometries are generated using the Neper software.\cite{2011-quey,2022-quey} The border-blocking growth rates are calculated using eqn~\eqref{eqn:growth-law-bb-exact} (thus correcting for film thickness variations between bubbles as in the 2D simulations). However, computational resources limit our 3D simulations to much smaller foam sizes than in 2D: in Fig.~\ref{fig:growth-law-3d}, we aggregate bubble data from four distinct $64$-bubble foams with periodic boundary conditions. While comparison to larger foams would be highly desirable to reduce finite-size effects and statistical fluctuations, Fig.~\ref{fig:growth-law-3d} does indicate fairly good agreement in the mean between the simulations and eqn~\eqref{eqn:growth-law-bb}, and gives some evidence to prefer the latter growth law over eqn~\eqref{eqn:growth-law-pasquet}, at least at higher $\phi$. However, we note that the level of agreement with eqn~\eqref{eqn:growth-law-bb} for small bubbles does worsen if the number of bins is doubled to $10$ (although the bin of smallest radius then contains few bubbles).
	
	As discussed later in Section~\ref{sec:discussion}, the scaling-state distribution is expected to be sensitive to the form of the growth law for small bubbles.\cite{2018-khakalo,2023-galvani} Fig.~\ref{fig:growth-law-2d} suggests that eqn~\eqref{eqn:growth-law-bb-2d} applies over a wide range of bubble sizes in 2D, although it would be desirable to test the growth law against simulations of more polydisperse foams, with more small bubbles. However, Fig.~\ref{fig:growth-law-3d} suggests that, in 3D, eqn~\eqref{eqn:growth-law-bb} may overestimate the shrinkage rate of small bubbles at higher $\phi$, perhaps due to the presence of rattlers; i.e.\ bubbles without contact films.\cite{2018-khakalo,2023-galvani} Further 3D simulations of larger systems may resolve this --- the discrepancy might be an artefact of aggregating data from several foams. The role of rattlers in a mean-field model remains unclear, and will be discussed in Section~\ref{sec:discussion}. In spite of these unresolved matters, the level of agreement in the mean observed in Figs.~\ref{fig:growth-law-2d} and~\ref{fig:growth-law-3d} (particularly the former, noting that the same arguments are used in 2D and 3D to derive the growth law) encourages us to explore the consequences of eqn~\eqref{eqn:growth-law-bb}: we derive its scaling state in Section~\ref{sub:reb:analysis}.
	
	First, however, we note that a foam is more commonly parametrised by liquid fraction $\phi$ than by osmotic pressure $\Pi_\text{O}$ (which appears in scaled form $\hat{\Pi}$ within the growth law). In a foam without bubble adhesion (such as we consider here), $\Pi_\text{O}$ is expected\cite{1986-princen,2021-hohler} to have a one-to-one correspondence with $\phi$ up to the unjamming transition $\phi_\text{c}$, beyond which $\Pi_\text{O} = 0$. We convert between $\phi$ and $\Pi_\text{O}$ using the empirical law\cite{2013-maestro,2021-hohler}
	\begin{align}
		\hat{\Pi} \approx (3.2) \, \frac{(\phi_\text{c} - \phi)^2}{(1 - \phi) \sqrt{\phi}} \label{eqn:osm-press}
	\end{align}
	for critical liquid fraction $\phi_\text{c} = 36\%$. The extra factor of $1-\phi$ compared to \citet{2013-maestro} is due to our different definition of the scaled osmotic pressure $\hat{\Pi}$ (above). In real foams, the critical liquid fraction depends on the bubble-size distribution, and is likely to be smaller in the polydisperse scaling-state distributions we derive below.\cite{2018-khakalo,2023-galvani} For example, \citet{2023-galvani} found $\phi_\text{c} = 31\%$ for a simulated foam with polydispersity equal to the foams in the ISS experiments. However, for concreteness, we select the standard value $\phi_\text{c} = 36\%$ which applies to monodisperse disordered 3D foams.\cite{2013-cantat}
	
	We also simplify the growth law~\eqref{eqn:growth-law-bb} by replacing all instances of $R_{3 2}$ by the critical radius $R_\text{c}$. Both measures of the average bubble size are expected to be fairly close in value. This approximation affects the derived scaling-state distributions very slightly (not shown), but gives the great advantage that the scaling-state properties can be derived analytically in closed form. This replacement is achieved by redefining $\hat{R} = R / R_\text{c}$ and $\hat{\Pi} = \Pi_\text{O} (R_\text{c} / \gamma) / (1 - \phi)$, and its very small effect on the predicted growth rates is shown in Fig.~\ref{fig:growth-law-3d}. We note that a similar replacement was made by \citet{1972-ardell} in a generalisation of the LSW law~\eqref{eqn:growth-law-lsw} to $\phi < 1$.
	
\section{Derivation of scaling state}\label{sub:reb:analysis}
	
	We now derive $\rho(\hat{R})$, the scaling-state bubble-size probability distribution predicted by the 3D border-blocking growth law~\eqref{eqn:growth-law-bb}. We follow the approach of \citeauthor{1983-marqusee},\cite{1983-marqusee} which is among the methods commonly used to derive scaling-state distributions in Ostwald ripening and coarsening theory.\cite{1959-lifshitz,1965-hillert,1977-binder,2006-streitenberger,2013-streitenberger} The derivation will also give us the growth exponent $\alpha$ (see Section~\ref{sec:intro}).
	
	Using a similar notation to \citeauthor{1983-marqusee},\cite{1983-marqusee} let $r = R / L$ and $\tau = t / T$ be dimensionless bubble radius and time variables, with $T = L^2 h / (2 \gamma D H)$ selected to simplify eqn~\eqref{eqn:growth-law-bb}, and with $L$ chosen such that the total (conserved) gas volume is unity.\footnote[3]{We define these variables partly for use in the simulations of Section~\ref{sec:sims}.} Let the dimensionless critical radius be $r_\text{c}(\tau) = R_\text{c}(\tau) / L$, which is expected to increase with time like the average bubble size\cite{2023-chieco} (with the same growth exponent\cite{2023-pasquet-b} $\alpha$) and define the dimensionless bubble radius $\hat{r} = r / r_\text{c} \, (= \hat{R})$ for consistency of notation. The growth law, eqn~\eqref{eqn:growth-law-bb}, becomes
	\begin{align}
	r'(r, \tau) \equiv \frac{d r}{d \tau} = \left(\frac{1}{r_\text{c}} - \frac{1}{r}\right) \frac{\hat{\Pi} \hat{r}}{1 + (1 + \hat{\Pi}) \hat{r}} . \label{eqn:growth-law-bb-dimless}
	\end{align}
	Let $n(r, \tau)$ be the bubble number distribution, such that $n(r, \tau) \, d r$ is the number of bubbles with radius between $r$ and $r + d r$ at time $\tau$, and $N(\tau) = \int_0^\infty n \, d r$ is the total number of bubbles. Since bubble radii vary continuously, the distribution $n$ satisfies the continuity equation\cite{1959-lifshitz}
	\begin{align}
	\dpart{n}{\tau} + \dpart{}{r} (r' n) = 0 . \label{eqn:continuity-eqn-n}
	\end{align}
	Following \citeauthor{1977-binder},\cite{1977-binder} let $x = r \tau^{-\alpha}$ be the scaled bubble size, recalling that $\alpha$ is the growth exponent (and so fixed $x$ corresponds to fixed $R / R_\text{c}$, for example). Define\cite{1959-lifshitz,1983-marqusee} $\bar{n}(x, \tau)$ as the number distribution in the variables $(x, \tau)$. Since $\bar{n} \, d x = n \, d r$, we have $\bar{n} = \tau^\alpha n$. Substituting this definition into eqn~\eqref{eqn:continuity-eqn-n}, and expressing in terms of the derivatives of the variables $(x, \tau)$ (noting that this change of variables alters $\partial / \partial \tau$) gives the continuity equation\cite{1959-lifshitz,1983-marqusee} for $\bar{n}$,
	\begin{align}
	\dpart{\bar{n}}{\tau} + \dpart{}{x} (u \bar{n}) = 0 , \label{eqn:continuity-eqn-nbar}
	\end{align}
	where\cite{1983-marqusee}
	\begin{align}
	u(x, \tau) = \frac{r'}{\tau^\alpha} - \frac{\alpha x}{\tau} \label{eqn:x-velocity}
	\end{align}
	is the velocity of a bubble's scaled radius $x$ (i.e.\ $d x / d \tau$).\cite{1959-lifshitz} We seek a scaling-state solution $\bar{n}(x, \tau)$ of eqn~\eqref{eqn:continuity-eqn-nbar}. Recalling Section~\ref{sec:intro}, this is a solution for which the distribution of bubble sizes is independent of time except for an overall scaling,\cite{2013-cantat} which is incorporated into the definition of $x$. Since bubbles shrink and disappear during coarsening,\cite{1999-weaire} the bubble number density $\bar{n}$ will decrease with time while the $x$ distribution remains constant. Hence, a scaling state takes the separable form\cite{1983-marqusee,2006-streitenberger} $\bar{n}(x, \tau) = \bar{\rho}(x) \bar{\sigma}(\tau)$, where $\bar{\rho}$ is proportional to the probability distribution in $x$ (which is $\tau$-independent in the scaling state), and $\bar{\sigma}$ accounts for gradual disappearance of bubbles, and thus the decay in the number density~$\bar{n}$.
	
	Following \citeauthor{1977-binder},\cite{1977-binder} we assume $\bar{\sigma} = \tau^{-\beta}$, and obtain the exponent $\beta$ using conservation of the amount of gas. \citet{1983-marqusee} account for the amount of dissolved discontinuous phase when imposing its conservation. However, we assume that, in foams, the equivalent volume of dissolved gas is negligible compared to the total bubble volume, presuming a sufficiently low gas solubility, and so take the total bubble volume to account for all gas in the system (which is conserved).\cite{1977-binder} We recall that $r$ is defined such that the total conserved gas volume is unity; i.e.\
	\begin{align}
	\frac{4 \pi}{3} \, \int_0^\infty r^3 n \, d r = 1 . \label{eqn:gas-conservation}
	\end{align}
	Substituting $r = x \tau^\alpha$, and using the fact that $n = \tau^{-\alpha} \bar{n} = \tau^{-\alpha - \beta} \bar{\rho}(x)$ in the scaling state, gives
	\begin{align}
	\frac{4 \pi}{3} \, \tau^{3 \alpha - \beta} \int_0^\infty x^3 \bar{\rho}(x) \, d x = 1 . \label{eqn:gas-conservation-x}
	\end{align}
	The right-hand side is evidently independent\cite{1977-binder} of time $\tau$, and so $\beta = 3 \alpha$ and $\bar{n} = \tau^{-3 \alpha} \bar{\rho}(x)$ in the scaling state. Substituting this form for $\bar{n}$ into eqn~\eqref{eqn:continuity-eqn-nbar}, we obtain\cite{1983-marqusee}
	\begin{align}
	\bar{\rho}(x) = \dpart{}{x} \left[\frac{\tau \, u(x, \tau)}{3 \alpha}
	\, \bar{\rho}(x)\right] . \label{eqn:dist-ode}
	\end{align}
	The exponent $\alpha$ is determined by noting that the right-hand side of this equation can have no explicit dependence\cite{1977-binder,1983-marqusee} on time $\tau$ (since the left-hand side has none), once eqns~\eqref{eqn:x-velocity} and~\eqref{eqn:growth-law-bb-dimless} are substituted for $u$. We define $x_\text{c} = r_\text{c} \tau^{-\alpha}$, which is $\tau$-independent in the scaling state since $\alpha$ is the growth exponent. Hence, $\hat{x} = x / x_\text{c} \, (= \hat{r})$ also has no explicit dependence on $\tau$ in the scaling state. Furthermore, $\hat{\Pi}$ is $\tau$-independent since we fix the liquid fraction $\phi$. By eqns~\eqref{eqn:x-velocity} and~\eqref{eqn:growth-law-bb-dimless},
	\begin{align}
	\tau u = \tau^{1-2\alpha} \left(\frac{1}{x_\text{c}} - \frac{1}{x}\right) \frac{\hat{\Pi} 
		\hat{x}}{1 + (1 + \hat{\Pi}) \hat{x}} - \alpha x . \label{eqn:x-velocity-scaled}
	\end{align}
	Substituting\cite{1977-binder,1983-marqusee} this result into eqn~\eqref{eqn:dist-ode} gives $\alpha = 1/2$, as expected in experiments using dry foams,\cite{1992-glazier,2010-lambert} and as in the small-$\phi$ model of \citeauthor{2017-schimming}.\cite{2017-schimming} However, eqn~\eqref{eqn:growth-law-bb} is intended as an approximate growth law for all $\phi$ at which bubbles are in contact, and for which $h$ is vanishingly small, so that gas transfer via films dominates.\cite{2017-schimming,2023-pasquet-b} Therefore eqn~\eqref{eqn:growth-law-bb} accords with the standard expectation that $\alpha = 1/2$ in any such border-blocking case.\cite{2023-pasquet-b} Substituting $\alpha = 1/2$ into eqn~\eqref{eqn:dist-ode}, and using that $\tau u$ is then $\tau$-independent, we obtain a linear, first-order ordinary differential equation for $\bar{\rho}$, which is straightforwardly solved (by separation of variables) to give\cite{1977-binder}
	\begin{align}
	\bar{\rho}(x) = -\frac{C x_\text{c}^3 / 2}{\tau u} \exp\left(\frac{3}{2} \int \frac{d x}{\tau u}\right) ,
	\label{eqn:dist-soln-1}
	\end{align}
	where $C$ is a constant. Rewriting eqn~\eqref{eqn:x-velocity-scaled}, for $\alpha = 1/2$, we obtain
	\begin{align}
	\tau u = -\frac{1}{2 x_\text{c}} \, \frac{(1 + \hat{\Pi}) x_\text{c} x^2
		+ (x_\text{c}^2 - 2 \hat{\Pi}) x + 2 \hat{\Pi} x_\text{c}}{x_\text{c}
		+ (1 + \hat{\Pi}) x} . \label{eqn:x-velocity-scaled-2}
	\end{align}
	The numerator is quadratic in $x$, so $\tau u$ has at most two zeros. We now follow standard arguments\cite{1959-lifshitz,1965-hillert,1983-marqusee,2006-streitenberger} to constrain these zeros, and thus to determine $x_\text{c}$ analytically.\footnote[4]{This is the reason for replacing $R_{3 2}$ by $R_\text{c}$ in eqn~\eqref{eqn:growth-law-bb}.\cite{1972-ardell} Otherwise, $x_\text{c}$ is obtained in terms of $x_{3 2} = R_{3 2} \tau^{-\alpha} / L$, which can only be obtained numerically.} First, we argue that $\bar{\rho}(x) = 0$ for all $x \geq x_0$, where $x_0 > 0$ is a finite cutoff.\cite{1959-lifshitz,1983-marqusee} Following \citeauthor{1983-marqusee},\cite{1983-marqusee} we note that, at large $x$, eqn~\eqref{eqn:dist-ode} becomes
	\begin{align}
	\frac{d \bar{\rho}}{d x} \approx -\frac{4 \bar{\rho}}{x} \label{eqn:dist-ode-large-x}
	\end{align}
	on substituting $\alpha = 1/2$ and eqn~\eqref{eqn:x-velocity-scaled}. Thus $\bar{\rho} = k / x^4$ for constant $k$. But $k = 0$ is the only value for which the integral in eqn~\eqref{eqn:gas-conservation-x} is finite.\cite{1983-marqusee} Therefore, $\bar{\rho}$ must be zero beyond some cutoff,\cite{1959-lifshitz,1983-marqusee}$^\text{,}$\footnote[5]{Linearity and homogeneity of eqn~\eqref{eqn:dist-ode} ensures that the argument would not be invalidated if higher order terms were included in eqn~\eqref{eqn:dist-ode-large-x}.} denoted $x_0$ (which we define to take its smallest possible value).
	
	Let us consider a bubble $b_0$ which initially has scaled radius $x = x_0$ in the scaling state, and which therefore is no smaller than any of the other bubbles. We recall\cite{1959-lifshitz} that $u$ gives the rate of change of $x$ for $b_0$. Thus, if $u < 0$ at $x_0$, then $x$ will decrease. But $u$ is a single-valued function of $x$ (and $\tau$) for all bubbles, by eqn~\eqref{eqn:x-velocity-scaled}, and so all bubbles initially with $x$ smaller than $x_0$ will remain smaller than $b_0$ as it shrinks.\cite{1959-lifshitz} Therefore, $\bar{\rho}$ will evolve,\cite{1965-hillert} with its cutoff equal to the decreasing scaled radius of $b_0$, contradicting the assumption that $\bar{\rho}$ is a function of $x$ only (and thus describes a scaling state). It follows that $u \geq 0$ at $x = x_0$.
	
	But $u$ should be nowhere positive for $x \leq x_0$ in the scaling state, because that would result in a class of bubbles which never vanish during coarsening,\cite{1965-hillert} contradicting the expected dynamics that all but one bubble should eventually vanish in a finite foam sample.\cite{1999-weaire,2013-cantat} Therefore, we must have\cite{1983-marqusee} $u = 0$ at $x = x_0$; i.e.\ the scaled radius of the bubble $b_0$ is constant.
	
	The quadratic numerator in eqn~\eqref{eqn:x-velocity-scaled-2} thus has a zero at $x = x_0$. From this equation, $u = -\hat{\Pi} / (\tau x_\text{c})$ at $x = 0$, and $u \sim -x / (2 \tau)$ as $x \to \infty$. We also recall that $\hat{\Pi} > 0$ in the jammed foams without bubble adhesion\cite{1979-princen,1986-princen} that we consider. Therefore, if $u$ has exactly a pair of distinct zeros, then both zeros are at $x > 0$, and $u > 0$ between these zeros. Hence, $x_0$ must equal the smaller of these zeros in such a case, since we have argued above that $u$ must be nonpositive for all $x \leq x_0$. But a stability problem then arises: if bubbles with $x$ just below the cutoff $x_0$ are perturbed to lie beyond $x_0$, then they will continue to grow\cite{1983-marqusee} (since $u > 0$), which will cause $\bar{\rho}$ to evolve. This is not a contradiction (since we have perturbed the distribution to begin with), but suggests that the scaling state is unstable in this case\cite{1983-marqusee,2022-li,2022-svoboda} (see the \hyperref[sec:app]{Appendix} for further discussion).
	
	Therefore, by the above exclusion of other possibilities, the desired stable scaling state should be such that $u$ has exactly one zero\cite{1959-lifshitz,1983-marqusee} at $x = x_0$; i.e.\ the quadratic in the numerator of eqn~\eqref{eqn:x-velocity-scaled-2} has a double zero at $x_0$. Hence, the quadratic's discriminant gives
	\begin{align}
	x_\text{c}^2 - 2 \hat{\Pi} = \pm 2 x_\text{c} \sqrt{2 \hat{\Pi} (1 + \hat{\Pi})} .
	\label{eqn:x-velocity-disc}
	\end{align}
	This quadratic equation can be solved for $x_\text{c}$, and the cutoff $x_0$ is then the double zero of the quadratic numerator in eqn~\eqref{eqn:x-velocity-scaled-2}. Therefore,\footnote[6]{Multiple roots arise, but only one has the required property that both $x_\text{c}$ and $x_0$ are nonnegative.}
	\begin{align}
	x_\text{c} = \sqrt{2 \hat{\Pi}} \left(\sqrt{2 + \hat{\Pi}}
	- \sqrt{1 + \hat{\Pi}}\right) , \quad x_0 = \sqrt{\frac{2 \hat{\Pi}}{1 + \hat{\Pi}}} . \label{eqn:x-crit-and-cutoff}
	\end{align}
	In the dry limit as $\hat{\Pi} \to \infty$, for which eqn~\eqref{eqn:growth-law-bb} reduces to Lemlich's law~\eqref{eqn:growth-law-lemlich},\cite{1961-wagner,1965-hillert,1978-lemlich} we have $x_\text{c} \to 1/\sqrt{2}$ and $x_0 \to \sqrt{2}$, in agreement with \citeauthor{1983-marqusee}.\cite{1983-marqusee} Furthermore, recalling that $x_0$ is the double zero of the numerator in eqn~\eqref{eqn:x-velocity-scaled-2}, we may rewrite this equation as
	\begin{align}
	\tau u = -\frac{1 + \hat{\Pi}}{2} \, \frac{(x_0 - x)^2}{x_\text{c}
		+ (1 + \hat{\Pi}) x} . \label{eqn:x-velocity-scaled-3}
	\end{align}
	To obtain the scaling-state distribution, we substitute this into eqn~\eqref{eqn:dist-soln-1}, giving
	\begin{align}
	\bar{\rho}(x) = C x_\text{c}^3 \ \frac{x + x_\text{c} / (1 + \hat{\Pi})}{(x_0 - x)^5}
	\, \exp\left[-3 \ \frac{x_0 + x_\text{c} / (1 + \hat{\Pi})}{x_0 - x}\right]
	\label{eqn:dist-soln-2}
	\end{align}
	for $x < x_0$. However, we desire the probability distribution $\rho(\hat{R})$ of relative bubble size in the scaling state. We recall that $\hat{R} \equiv R / R_\text{c} = x / x_\text{c}$ by definition (which has no explicit $\tau$-dependence in the scaling state), while $\rho(\hat{R}) \, d \hat{R} = \bar{\rho}(x) \, d x$. Thus, $\rho(\hat{R}) = x_\text{c} \bar{\rho}(x)$. We also define the dimensional cutoff radius $R_0 = x_0 L \tau^\alpha$ with $\alpha = 1/2$, analogous to $R$ and $R_\text{c}$, and its relative value $\hat{R}_0 = R_0 / R_\text{c} \, (= x_0 / x_\text{c})$. Therefore, eqn~\eqref{eqn:dist-soln-2} gives
	\begin{align}
	\rho(\hat{R}) = C \ \frac{\hat{R} + 1 / (1 + \hat{\Pi})}{(\hat{R}_0 - \hat{R})^5}
	\, \exp\left[-3 \, \frac{\hat{R}_0 + 1 / (1 + \hat{\Pi})}{\hat{R}_0 - \hat{R}}\right]
	\label{eqn:probability-dist}
	\end{align}
	as the probability distribution of relative bubble size in the scaling state, where $C$ is chosen to normalise the distribution. Eqn~\eqref{eqn:probability-dist} is valid for $\hat{R} < \hat{R}_0$, whereas $\rho(\hat{R}) = 0$ otherwise. From eqn~\eqref{eqn:x-crit-and-cutoff}, the relative cutoff is
	\begin{align}
	\hat{R}_0 = 1 + \sqrt{\frac{2 + \hat{\Pi}}{1 + \hat{\Pi}}} . \label{eqn:r-cutoff}
	\end{align}
	In the dry limit $\hat{\Pi} \to \infty$, the distribution becomes
	\begin{align}
	\rho(\hat{R}) = C \ \frac{\hat{R}}{(\hat{R}_0 - \hat{R})^5}
	\, \exp\left[-\frac{3 \hat{R}_0}{\hat{R}_0 - \hat{R}}\right] , \ \text{with $\hat{R}_0 = 2$} , \label{eqn:probability-dist-dry}
	\end{align}
	which is equivalent to the scaling-state distribution\cite{1961-wagner,1965-hillert,1985-markworth} of Lemlich's law~\eqref{eqn:growth-law-lemlich}, to which eqn~\eqref{eqn:growth-law-bb} reduces in this limit. In the limit $\hat{\Pi} \to 0$, corresponding to the approach to the unjamming transition, eqn~\eqref{eqn:probability-dist} becomes
	\begin{align}
	\rho(\hat{R}) = C \ \frac{1 + \hat{R}}{(\hat{R}_0 - \hat{R})^5}
	\, \exp\left[-3 \, \frac{1 + \hat{R}_0}{\hat{R}_0 - \hat{R}}\right] , \ \text{with $\hat{R}_0 = 1 + \sqrt{2}$} . \label{eqn:probability-dist-wet}
	\end{align}
	We recall that gas transfer through the bulk liquid is neglected in our model, because we take $h \to 0$. Hence, eqn~\eqref{eqn:probability-dist-wet} differs from the scaling-state distribution\cite{1959-lifshitz} predicted by the LSW law~\eqref{eqn:growth-law-lsw}, for which all gas diffusion is via bulk liquid rather than films.\cite{2013-cantat}
	
\begin{figure}[h]
		\centering
		\includegraphics[width=\columnwidth]{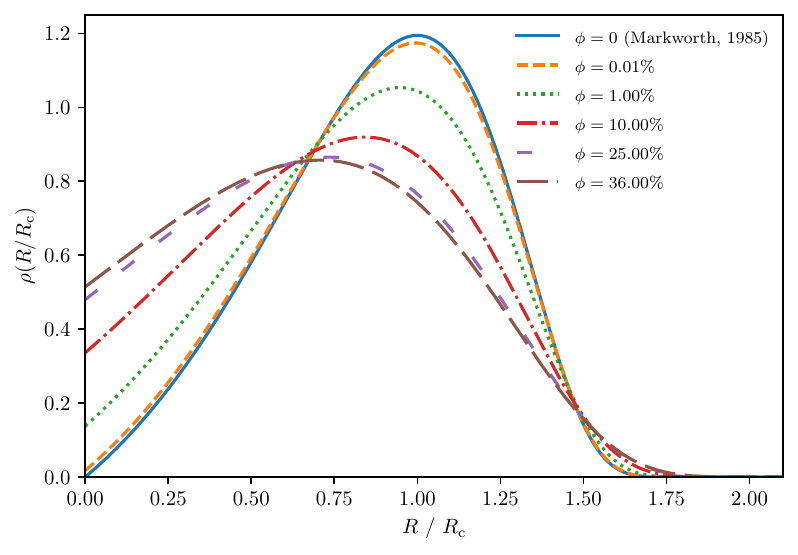}
		\caption{Probability distribution $\rho$ of the relative bubble size $\hat{R}$ in the scaling state, predicted by eqn~\eqref{eqn:probability-dist} ($C$ is obtained by numerical integration of the distribution), for various liquid fractions. We also plot $\rho$ in the limiting cases of a dry foam and a foam at the unjamming transition, given respectively by eqns~\eqref{eqn:probability-dist-dry}\cite{1961-wagner,1985-markworth} and~\eqref{eqn:probability-dist-wet}.}
		\label{fig:scaling-state-dists}
	\end{figure}

	We plot eqn~\eqref{eqn:probability-dist} in Fig.~\ref{fig:scaling-state-dists} for a variety of liquid fractions $\phi$, which determine $\hat{\Pi}$ via eqn~\eqref{eqn:osm-press}.\cite{2013-maestro} We also include the above limiting distributions.
	
	Perhaps the most noticeable property of the probability distribution~\eqref{eqn:probability-dist} is that, for all wet foams, we have $\rho > 0$ at $R = 0$, which is a qualitative difference from its form in the dry limit~\eqref{eqn:probability-dist-dry}.\cite{1965-hillert,1985-markworth} This may be interpreted using the approximation~\eqref{eqn:film-area-approx} for the relative film area $A / S$ of bubbles: smaller bubbles have a smaller proportion of their surface in contact with neighbouring bubbles,\cite{1991-bolton} and thus relatively less area available for gas transfer under the assumption of border blocking.\cite{1991-bolton,2013-roth} In our mean-field model, $A / S$ tends to zero as $R \to 0$; i.e.\ bubbles of zero size are rattlers. Therefore, in spite of the divergence of bubble pressure as $R \to 0$, a large population of small bubbles remains, in the scaling state, due to their disproportionately small films.\footnote[7]{However, integrating eqn~\eqref{eqn:growth-law-bb} for small $R$ shows that a small bubble does vanish in finite time. Although the film area goes to zero as $R \to 0$, the bubble pressure diverges, and so $d R / d t$ tends to a finite limit.} In the dry limit,\cite{1978-lemlich} $A / S = 1$ independent of $R$, and so this effect does not arise. We discuss the predicted probability distribution~\eqref{eqn:probability-dist} further in Section~\ref{sec:discussion}, where we compare it to prior simulations and experiments.\cite{2018-khakalo,2023-galvani,2025-galvani} For now, we note that a large population of small bubbles was also found in these studies, but the form of the bubble size distribution is qualitatively different.\cite{2017-khakalo-pre-n,2018-khakalo,2023-galvani,2025-galvani}
	
\begin{figure}[h]
		\centering
		\includegraphics[width=\columnwidth]{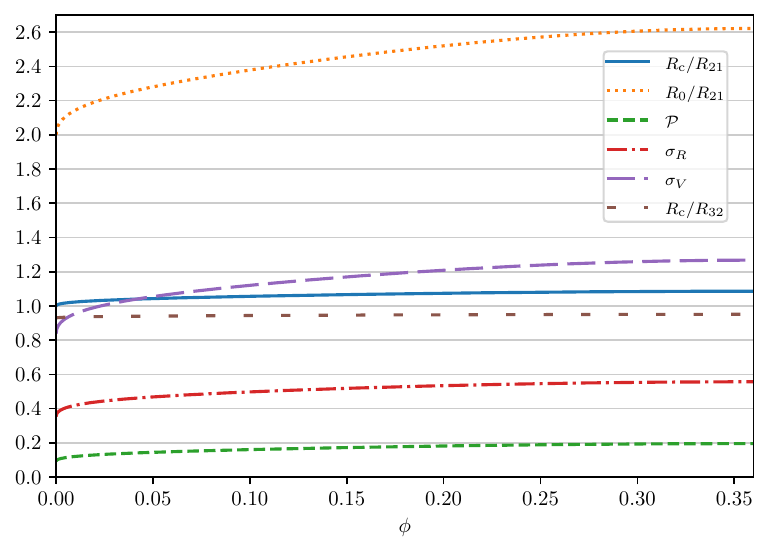}
		\caption{Various properties of eqn~\eqref{eqn:probability-dist} as functions of liquid fraction. The ratios of critical radius and cutoff radius to $R_{2 1}$ (and $R_{3 2}$ in the former case) are shown, alongside the polydispersity\cite{2004-kraynik} $\mathcal{P}$ and geometric disorder\cite{2013-cantat} $\sigma_V$, which are defined in the text, and the standard deviation $\sigma_R$ of $R / \langle R \rangle$. These are all evaluated numerically from eqns~\eqref{eqn:probability-dist} and~\eqref{eqn:r-cutoff}.}
		\label{fig:dist-props}
	\end{figure}

	In Fig.~\ref{fig:dist-props}, we give the variation with $\phi$ of $R_\text{c} / R_{2 1}$, $R_\text{c} / R_{3 2}$, $R_0 / R_{2 1}$, polydispersity\cite{2004-kraynik} $\mathcal{P} = R_{3 2} / \langle R^3 \rangle^{1/3} - 1$, the standard deviation $\sigma_R$ of $R / \langle R \rangle$, and geometric disorder\cite{2013-cantat} $\sigma_V = \sqrt{\langle V^2 \rangle / \langle V \rangle^2 - 1}$ in the scaling state (where $V = 4 \pi R^3 / 3$ is the bubble volume). We recall from Section~\ref{sec:intro} that $R_\text{c} = R_{2 1}$ in the dry limit.\cite{1978-lemlich} All quantities plotted in Fig.~\ref{fig:dist-props} are measures of the distribution's width, and so they each increase with $\phi$, consistent with the broadening of the distribution apparent from Fig.~\ref{fig:scaling-state-dists}. In Fig.~\ref{fig:dist-props}, $R_\text{c} / R_{3 2} \approx 1$, which suggests that our replacement of $R_{3 2}$ by $R_\text{c}$ in eqn~\eqref{eqn:growth-law-bb} (see Section~\ref{sec:growth-law-bb}) is self consistent. The poor agreement of the polydispersity in Fig.~\ref{fig:dist-props} with the ISS experiments (for which $\mathcal{P} > 0.3$ for $\phi \lesssim 30\%$, and $\mathcal{P}$ tends to decrease as $\phi$ increases\cite{2023-galvani}) is interpreted in Section~\ref{sec:discussion}.
	
\begin{figure}[h]
		\centering
		\includegraphics[width=\columnwidth]{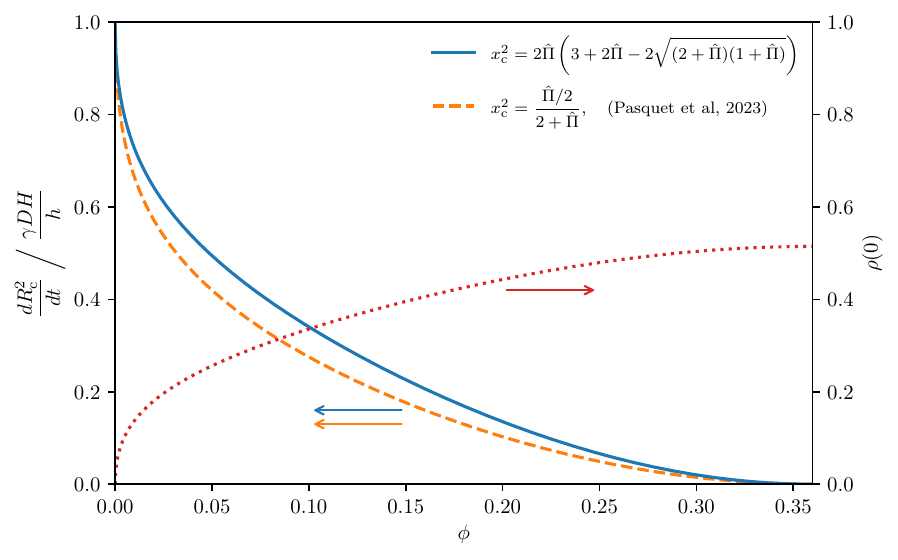}
		\caption{Coarsening rate (see text) versus liquid fraction $\phi$, as predicted by the growth law~\eqref{eqn:growth-law-bb} via eqn~\eqref{eqn:x-crit-and-cutoff}, alongside the prediction of \citeauthor{2023-pasquet-b}\cite{2023-pasquet-b} The predictions agree at $\phi = 0$, where they equal unity with the chosen scaling, and they both reach zero at $\phi = \phi_\text{c}$. The variation of $\rho(0)$ with $\phi$, given by eqn~\eqref{eqn:probability-dist} (with $C$ again being obtained numerically), is also plotted ($\rho(0) = 0$ at $\phi = 0$).}
		\label{fig:coarsen-rate}
	\end{figure}

	We recall, by the definition of $x_\text{c}$, that the critical radius satisfies $r_\text{c} = x_\text{c} \tau^\alpha$ in the scaling state, where $\alpha = 1/2$. But $x_\text{c}$ is a constant, and therefore so is $d r_\text{c}^2 / d \tau = x_\text{c}^2$, which provides a measure of the coarsening rate of the foam\cite{2023-pasquet-b,2025-requier} (proportional to the rate at which a typical bubble's surface area increases). We plot this in Fig.~\ref{fig:coarsen-rate} as a function of $\phi$, and compare it with the corresponding rate predicted by \citet{2023-pasquet-b} using the simpler growth law~\eqref{eqn:growth-law-pasquet}.
	
	As expected, the coarsening rate decreases with increasing $\phi$ in Fig.~\ref{fig:coarsen-rate} due to the decrease in film areas, reaching zero when the bubbles lose contact at the unjamming transition. Our prediction is close to that of \citet{2023-pasquet-b} (agreeing at $\phi = 0$ since the two growth laws are equivalent in this case), although our rates are slightly higher for intermediate $\phi$. This is perhaps counterintuitive, as our model has enhanced border-blocking for small bubbles compared to eqn~\eqref{eqn:growth-law-pasquet}, thus slowing their disappearance along, presumably, with the growth of the average bubble size. However, as expressed in more detail in Section~\ref{sec:discussion}, the rate of change of the bubble number $d N / d t$ (and thus the coarsening rate) depends on $\rho(0)$ in addition to the limiting growth rate as $R \to 0$.\cite{1986-mullins,2012-lambert}
	
	We note that the prediction\cite{2023-pasquet-b} $x_\text{c}^2 = \hat{\Pi} / (4 + 2 \hat{\Pi})$ plotted in Fig.~\ref{fig:coarsen-rate}, which originates from the film area approximation~\eqref{eqn:film-area-hohler},\cite{2021-hohler} has been successfully fitted\cite{2025-galvani} to data from coarsening experiments at small $\phi \leq 8\%$. However, its agreement with the ISS experiments at larger $\phi$ is poor, likely due to bubble adhesion.\cite{2023-pasquet-b,2025-requier} Recalling that we consider only nonadhesive foams in the present work, we consider this discrepancy no further, except to note that an adaptation of eqn~\eqref{eqn:film-area-hohler} to adhesive foams has been proposed and used by \citet{2025-galvani}
	
	Having derived and analysed the scaling-state bubble size distribution~\eqref{eqn:probability-dist} predicted by the border-blocking mean-field growth law~\eqref{eqn:growth-law-bb}, we next give results from mean-field simulations to verify that this distribution is approached from a range of initial conditions.\cite{1983-marqusee,2022-li}

\section{Simulations}\label{sec:sims}

\subsection{Simulation methods}\label{sec:methods}
	
	We now describe the approach we use to simulate foams that coarsen according to a mean-field bubble growth law of the LSW\cite{1959-lifshitz,1961-wagner} or Lemlich\cite{1961-wagner,1965-hillert,1978-lemlich} type; i.e.\ which depends only on bubble radius $R$, like eqn~\eqref{eqn:growth-law-bb}.
	
	The numerical methods are adapted from those of \citeauthor{1997-de-smet},\cite{1997-de-smet} and hence involve predicting the evolution of a large number of individual bubbles (rather than calculating the evolution of a distribution function\cite{1978-lemlich,1997-de-smet,2022-li}). We select this approach for its directness, and we note that present computational resources allow considerably larger system sizes to be studied than were simulated by \citeauthor{1997-de-smet}\cite{1997-de-smet} To our understanding, a similar approach is used by \citeauthor{1989-brown}.\cite{1989-brown} We implement the simulations in \emph{python}, using the \emph{numpy} and \emph{scipy} packages.\cite{2023-numpy,2023-scipy}
	
	We use the dimensionless bubble radius and time variables $r$ and $\tau$ defined in Section~\ref{sub:reb:analysis}. Let $v = 4 \pi r^3 / 3$ be the dimensionless volume of a bubble.
	
	We begin the simulation with $N = 3 \times 10^7$ bubbles with different radii $r_i$ (and volumes $v_i$). The initial bubble number $N$ is chosen to be as large as is feasible (we use a PC with a $16$-core Intel i7 processor from circa 2021, $\unit[16]{GB}$ of RAM, and an SSD), noting that memory usage appears to be the limiting factor. We make no attempt to parallelise the code. For comparison, \citet{2006-thomas} use initial foams containing about $2 \times 10^6$ bubbles in their 3D Potts-model simulations of dry-foam coarsening (which account for the detailed foam structure, unlike our mean-field simulations). We sample the relative radii $r_i / \langle r \rangle$ from a specified distribution, usually a narrow\cite{2022-li} lognormal distribution with standard deviation $\sigma_R = 1/10$. The radii $r_i$ themselves are determined by imposing the condition (see Section~\ref{sub:reb:analysis}) that the total gas volume is $\sum_i v_i = \sum_i 4 \pi r_i^3 / 3 = 1$. We select a lognormal distribution since this is frequently\cite{1965-hillert,2010-lambert,2018-quey} used to approximate the scaling-state distribution of three-dimensional coarsening systems with $\phi = 0$, although as an initial condition the choice is fairly arbitrary. See Section~\ref{sub:reb:sims} for a discussion of the above choice of $\sigma_R$, along with the other initial distributions we use.
	
	Next, the main loop is entered, each pass through which evolves the foam by a single time step $\Delta \tau$ (to be defined below). At the beginning of each step, the critical radius $r_\text{c}$ is calculated numerically by solving $\langle d v / d \tau \rangle = 0$; i.e.\ by imposing that the total volume of gas is conserved.\cite{1978-lemlich,2013-cantat}
	
	Once $r_\text{c}$ is obtained, the growth rate $d r / d \tau$ or $d v / d \tau$ of each bubble is known from eqn~\eqref{eqn:growth-law-bb-dimless}. The root-mean-square volumetric growth rate $\sqrt{\langle (d v / d \tau)^2 \rangle}$ is used to set the time step $\Delta \tau$, which we define such that $m = 10^3$ steps (this parameter has been varied to check convergence; see the \hyperref[sec:app]{Appendix}) would be required for a hypothetical bubble initially with the mean volume $\langle v \rangle$ to shrink to zero size if its growth rate were fixed at the current value of $-\sqrt{\langle (d v / d \tau)^2 \rangle}$. We use this adaptive definition of $\Delta \tau$ for numerical efficiency, noting that the rate of coarsening (measured by $d r_\text{c} / d \tau$ for example) is expected to slow with time in a real foam (since $r_\text{c} \sim \tau^\alpha$ for $\alpha < 1$ in the scaling state;\cite{2013-isert,2023-pasquet-b} in our model, $\alpha = 1/2$ by Section~\ref{sub:reb:analysis}).
	
	Statistics of the foam, including the mean bubble radius $\langle r \rangle$ and polydispersity\cite{2004-kraynik} $\mathcal{P}$, are then calculated and output. Every $300$ time steps, we store the bubble radius distribution as a histogram in $r / r_\text{c}$, with $100$ equal bins between $r = 0$ and $r/r_\text{c} = 2.5$, which we find sufficient to characterise the distribution (indeed, we undersample the histogram in Section~\ref{sub:reb:sims}).
	
	A single Euler step\cite{2013-adams} is then applied to evolve the bubble volumes; i.e.\ $v_i(\tau + \Delta \tau) = v_i(\tau) + (d v_i / d \tau) \, \Delta \tau$. More efficient algorithms might allow fewer (larger) time steps to be used without increasing error in the bubble radii. However, the execution time of a simulation is fairly short (requiring about $4$~hours on the PC described above). Furthermore, we believe that the main inaccuracy induced by the finite time steps is due to the size threshold for small bubbles, which we now discuss.
	
	We recall that small bubbles shrink and eventually vanish due to coarsening.\cite{1978-lemlich} In the simulations, we define a minimum bubble volume allowed following the above Euler step, below which a bubble is deleted. In order to avoid bubbles of negative volume, we delete a bubble if its volume $v$ is small enough that the bubble would vanish after two further time steps $\Delta \tau$ at its present growth rate $d v / d \tau$ (both $\Delta \tau$ and $d v / d \tau$ are fixed at their values calculated prior to the Euler step). The choice of two steps is arbitrary, but is intended to allow for the variation of $\Delta \tau$ in future steps (recalling that this is chosen adaptively). The total volume of all the deleted bubbles is reassigned to the bubbles that remain by multiplying their volumes by a constant factor such that the total remains unity.\cite{1978-lemlich,2000-gardiner} The minimum bubble volume affects the bubble-size distribution at small radii, as discussed in the \hyperref[sec:app]{Appendix}.
	
	This completes the time step, following which the next pass of the main loop is performed. We continue until fewer than $10^3$ bubbles remain, at which point the simulation is halted since we expect smaller systems to exhibit excessive statistical fluctuations\cite{2006-thomas} (see Figs.~\ref{fig:crit-rad-diff-v-time} and~\ref{fig:poly-v-rad} below).

\subsection{Simulation results}\label{sub:reb:sims}
	
	Using the methods\cite{1997-de-smet} described in Section~\ref{sec:methods}, we perform coarsening simulations to first verify that the distribution~\eqref{eqn:probability-dist} is reproducible at a variety of liquid fractions $\phi$ for a given initial bubble-size distribution. We then vary the initial distribution to provide evidence that the derived scaling state is universal, as it is expected to be in a real foam.\cite{2013-cantat} Finally, we compare the evolution of other foam properties (such as the critical radius $R_\text{c}$) with theory. In the \hyperref[sec:app]{Appendix}, we describe exceptions to the universality of the scaling state,\cite{1989-brown,2022-li} and discuss convergence of our simulations with respect to the time-step size.
	
	We emphasise that the simulations use the same growth law~\eqref{eqn:growth-law-bb} as analysed in Section~\ref{sub:reb:analysis}. Hence, the growth law itself and the modelling assumptions are untested by the simulation results. Rather, our aim here is to show that the derived scaling state is approached at late times over a range of $\phi$ and from a variety of initial conditions (which is not proven by the derivation).
	
\begin{figure}[h!]
		\centering
		\includegraphics[width=\columnwidth]{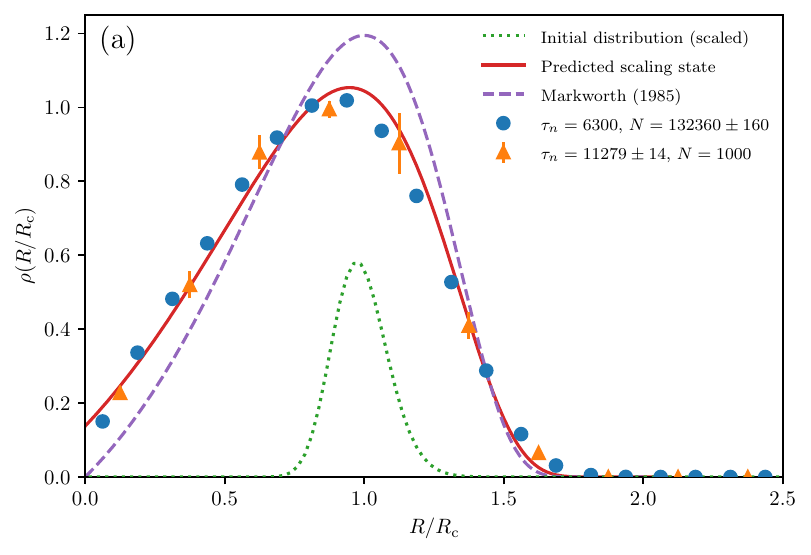}
		\includegraphics[width=\columnwidth]{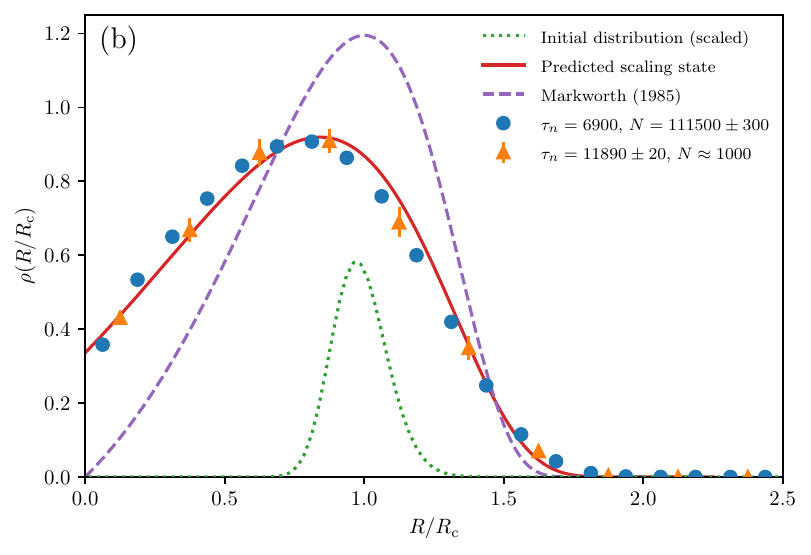}
		\includegraphics[width=\columnwidth]{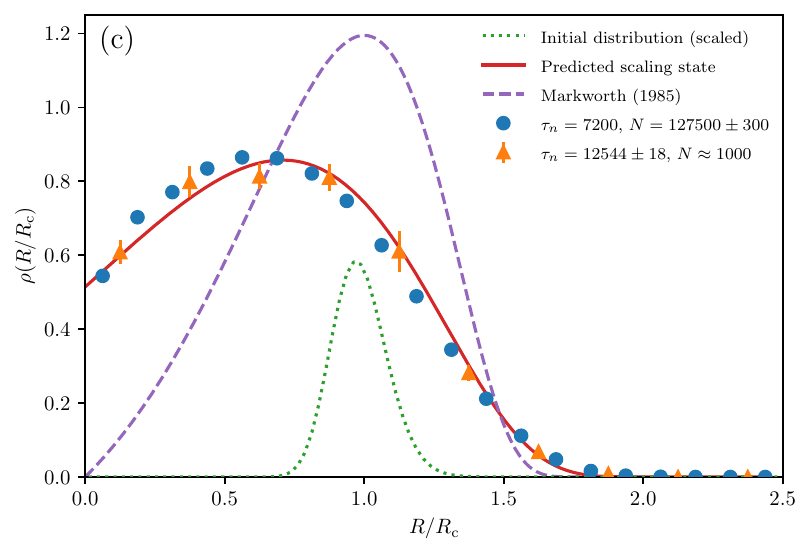}
		\caption{Simulated bubble-size histograms for (a) $\phi = 1\%$, (b) $\phi = 10\%$, and (c) $\phi = 35\%$. The mean over five simulation runs is shown, and error bars give the sample standard deviation (where larger than the marker) over the five runs. The distributions are plotted when roughly $N = 10^5$ bubbles remain (with $20$ bins), and when the runs halt (with $10$ bins). The corresponding number $\tau_n$ of completed time steps is given. The final $N$ value is not exactly $10^3$ because multiple bubbles can disappear in one time step. The uncertainties in the legend are due to differences between the five runs. Comparison is made to the predicted scaling state~\eqref{eqn:probability-dist}, and to that for the dry limit, eqn~\eqref{eqn:probability-dist-dry}.\cite{1961-wagner,1965-hillert,1985-markworth} The initial distribution (stated in the text) is also shown (not normalised, for clarity).}
		\label{fig:sim-dist-v-liq-frac}
	\end{figure}

	In Fig.~\ref{fig:sim-dist-v-liq-frac}, we plot bubble-size histograms from our simulations for $\phi \in \{1\%, 10\%, 35\%\}$. The initial bubble sizes are taken from a narrow lognormal distribution with standard deviation $\sigma_R = 1/10$ with respect to the relative radius $R / \langle R \rangle$. Data from five simulations is aggregated in each case (with distinct samples from the same initial distribution), and we plot histograms when about $10^5$ bubbles remain and when about $10^3$ bubbles remain (i.e.\ at the end of the simulations). We observe from Fig.~\ref{fig:sim-dist-v-liq-frac} that agreement with eqn~\eqref{eqn:probability-dist} is good at each liquid fraction (whereas the initial distribution is dissimilar from the predicted scaling state), and the nonzero value of $\rho(\hat{R} = 0)$ is reproduced to a fair approximation.
	
\begin{figure}[h!]
		\centering
		\includegraphics[width=\columnwidth]{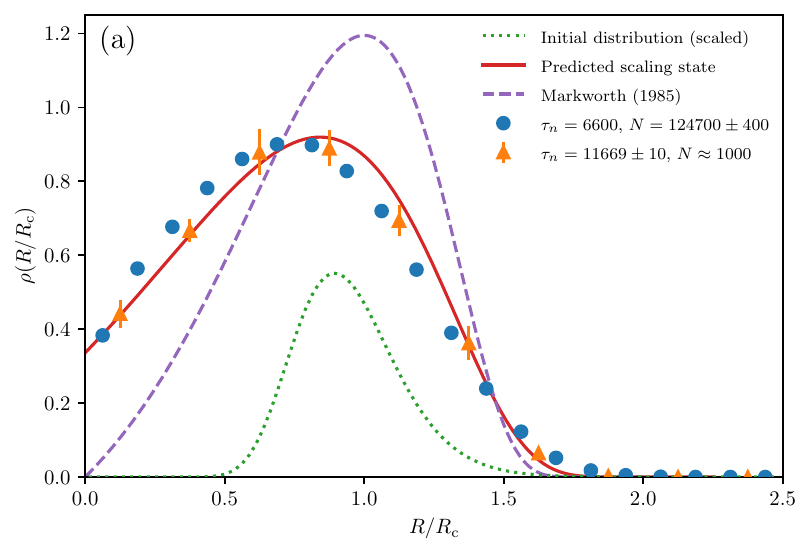}
		\includegraphics[width=\columnwidth]{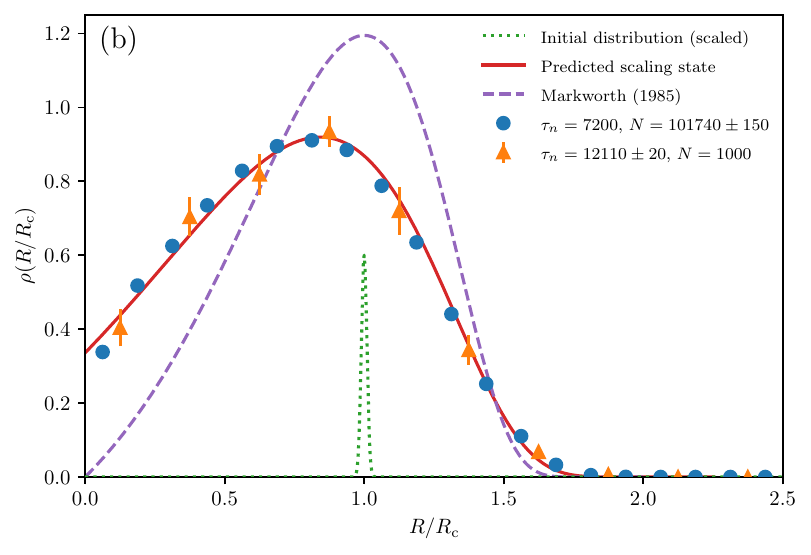}
		\includegraphics[width=\columnwidth]{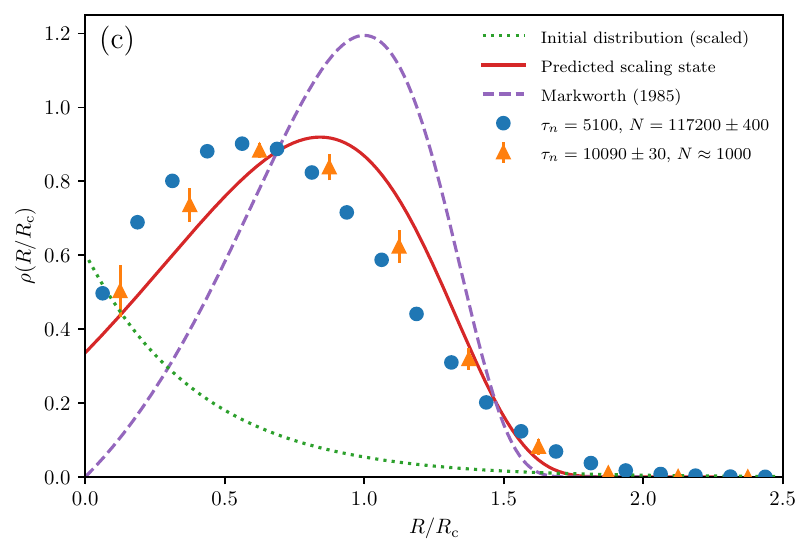}
		\caption{Simulated bubble-size histograms for $\phi = 10\%$. These plots differ from Fig.~\ref{fig:sim-dist-v-liq-frac}(b) only in the initial bubble-size distributions (which are again shown unnormalised for clarity). In (a), the initial distribution is lognormal with standard deviation $\sigma_R = 1/5$ in $R / \langle R \rangle$. The distribution is lognormal with $\sigma_R = 1/100$ in (b), and is exponential in (c).}
		\label{fig:sim-dist-v-init-dist}
	\end{figure}

	Therefore, the derived scaling state is approached from one initial condition for a range of liquid fractions. In Fig.~\ref{fig:sim-dist-v-init-dist}, we vary the initial bubble-size distribution (defined in the caption) to show that the theoretical prediction is approached from a variety of initial conditions (similar plots are given by \citet{2022-li} for a different growth law). Comparison should be made with Fig.~\ref{fig:sim-dist-v-liq-frac}(b). We note that the speed with which the system approaches the scaling state (judged by eye using the deviation from theory at the end of the simulations) varies with the initial distribution, and we see that wider distributions tend to result in a longer transient, as observed in other coarsening simulations with different growth laws (though exceptions have previously been observed for extremely narrow initial distributions).\cite{1997-de-smet,2022-li} This is the reason we select a narrow initial distribution in Section~\ref{sec:methods} (lognormal with standard deviation $\sigma_R = 1/10$) for use in most of our simulations.
	
\begin{figure}[h]
		\centering
		\includegraphics[width=\columnwidth]{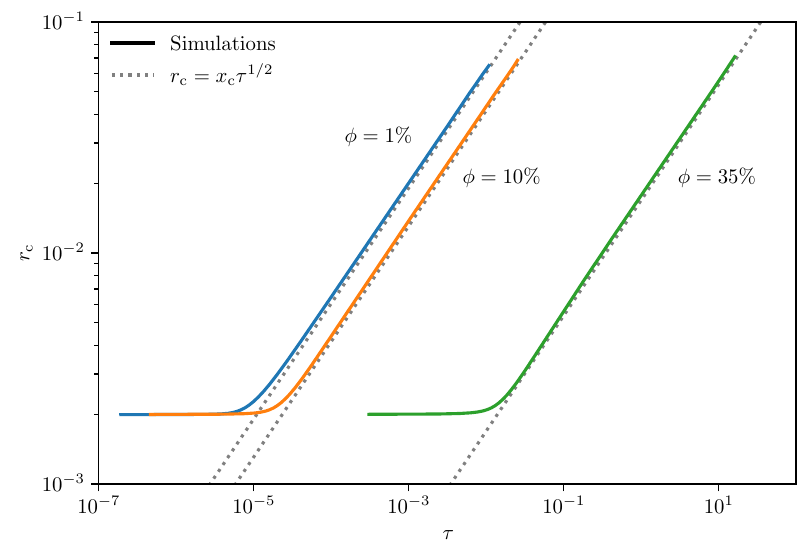}
		\caption{Dimensionless critical radius $r_\text{c}$ versus dimensionless time $\tau$, on logarithmic scales, from simulations at different liquid fractions $\phi$. Comparison is made to the scaling-state derivation of Section~\ref{sub:reb:analysis}, where $x_\text{c}$ is from eqn~\eqref{eqn:x-crit-and-cutoff}, as a test of the derivation itself (not of the modelling assumptions). The initial bubble-size distributions are lognormal with $\sigma_R = 1/10$, as in Fig.~\ref{fig:sim-dist-v-liq-frac}. The simulation curves start after ten time steps, to ensure they are sufficiently resolved.}
		\label{fig:crit-rad-v-time}
	\end{figure}

\begin{figure}[h]
		\centering
		\includegraphics[width=\columnwidth]{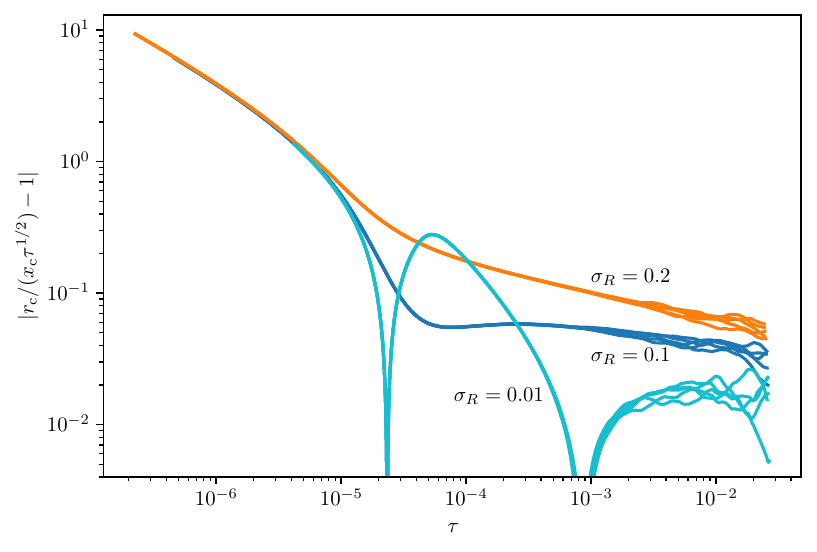}
		\caption{Relative difference between the simulated critical radius $r_\text{c}$ and the scaling-state prediction $x_\text{c} \tau^{1/2}$, versus time. Eqn~\eqref{eqn:x-crit-and-cutoff} gives $x_\text{c}$, and the liquid fraction is $\phi = 10\%$. Runs are plotted for lognormal initial bubble-size distributions with various standard deviations $\sigma_R$; five independent runs are plotted per value of $\sigma_R$, which overlap until late times. The curves start after ten time steps, as in Fig.~\ref{fig:crit-rad-v-time}.}
		\label{fig:crit-rad-diff-v-time}
	\end{figure}

	Having shown that the scaling-state distribution~\eqref{eqn:probability-dist} is approached from a variety of initial conditions, we now plot the time-dependence of a few foam statistics as a further test of the derivation in Section~\ref{sub:reb:analysis} (again, we emphasise that this is not a test of the modelling assumptions). We start with the evolution of critical radius $r_\text{c}$ for a variety of liquid fractions, in Fig.~\ref{fig:crit-rad-v-time}.\cite{2013-isert,2018-khakalo,2023-pasquet-b} Good agreement is observed with the scaling-state prediction $x_\text{c} \tau^{1/2}$ after a transient of the expected qualitative form;\cite{2000-gardiner,2023-chieco} where $x_\text{c}$ is given by eqn~\eqref{eqn:x-crit-and-cutoff}. The offset of the curves corresponds to the decrease in coarsening rate with liquid fraction (c.f.\ Fig.~2 of \citet{2023-pasquet-b}). The relative difference of $r_\text{c}$ from $x_\text{c} \tau^{1/2}$ is plotted in Fig.~\ref{fig:crit-rad-diff-v-time} for various initial conditions at $\phi = 10\%$ to reinforce this agreement. All plotted runs exhibit a relative difference of below $0.1$ at their termination. Consistent with Fig.~\ref{fig:sim-dist-v-init-dist}, narrower initial distributions tend to approach the theoretical scaling state more closely by the end of the simulations;\cite{1997-de-smet,2022-li} i.e.\ they typically exhibit a smaller deviation between $r_\text{c}$ and $x_\text{c} \tau^{1/2}$. The two cusps in the curves for the narrowest initial condition correspond to changes in the sign of $r_\text{c} - x_\text{c} \tau^{1/2}$: very narrow initial conditions can cause $r_\text{c}$ to undershoot its scaling-state prediction, consistent with the theory of \citeauthor{2023-chieco}.\cite{2023-chieco}
	
\begin{figure}[h]
		\centering
		\includegraphics[width=\columnwidth]{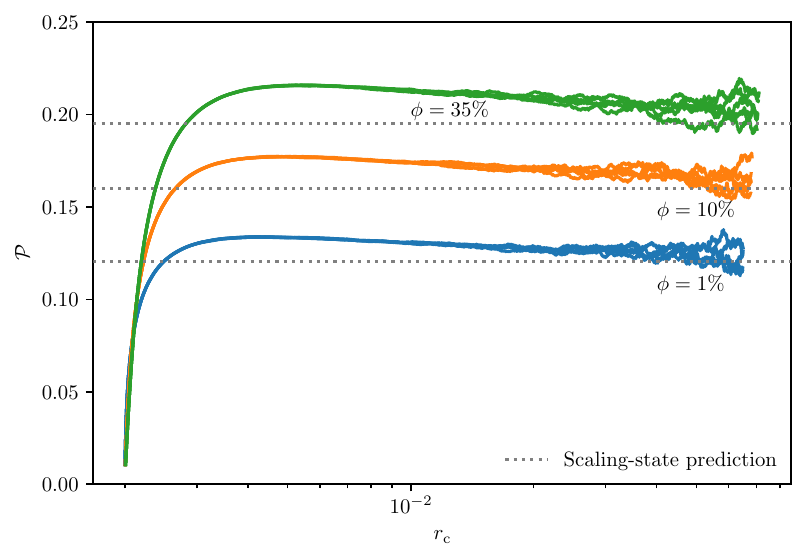}
		\caption{Polydispersity\cite{2004-kraynik} $\mathcal{P}$ versus dimensionless critical radius, for various liquid fractions $\phi$. As in Figs.~\ref{fig:sim-dist-v-liq-frac} and~\ref{fig:crit-rad-v-time}, the initial bubble-size distribution is lognormal with $\sigma_R = 1/10$. Five runs per $\phi$ are plotted, and the curves start after ten time steps.}
		\label{fig:poly-v-rad}
	\end{figure}

	Finally, Fig.~\ref{fig:poly-v-rad} compares the polydispersity\cite{2004-kraynik} $\mathcal{P}$ (see Section~\ref{sub:reb:analysis}) with its predicted scaling-state value for several liquid fractions. The critical radius $r_\text{c}$ is used as the independent variable here, to avoid an offset between the curves. We see that the simulations tend towards the theoretical predictions, which are obtained via numerical integration of eqn~\eqref{eqn:probability-dist}. The approach to these scaling-state values depends on the initial conditions (data not shown).
	
	In Figs.~\ref{fig:crit-rad-diff-v-time} and~\ref{fig:poly-v-rad}, we have plotted five simulation runs per liquid fraction (differing in the initial bubble-size samples), since this conveniently indicates, by the degree to which the curves `fray,' the importance of statistical fluctuations at a given time.
	
	The results of this section show that the scaling state derived in Section~\ref{sub:reb:analysis} is approached from a range of initial conditions (noting the exceptions\cite{1989-brown,2022-li,2022-svoboda} discussed in the \hyperref[sec:app]{Appendix}) in simulations which apply the mean-field growth law~\eqref{eqn:growth-law-bb} directly. This validates the derivation of Section~\ref{sub:reb:analysis} and indicates that the derived scaling state is stable.

\section{Discussion}\label{sec:discussion}
	
	The scaling-state bubble-size distribution~\eqref{eqn:probability-dist} derived from the mean-field growth law~\eqref{eqn:growth-law-bb} is qualitatively different for $\phi > 0$ from the distribution~\eqref{eqn:probability-dist-dry} predicted by Lemlich's law. The distribution $\rho(R)$ described by eqn~\eqref{eqn:probability-dist} is nonzero at $R = 0$ for $\phi > 0$,  whereas $\rho(0) = 0$ for eqn~\eqref{eqn:probability-dist-dry}. To show this directly, we use the following result (which we have expressed in terms of $R$): the relative rate at which bubbles disappear is\cite{1986-mullins,2012-lambert}
	\begin{align}
		 \frac{1}{N} \frac{d N}{d t} = \lim_{R \to 0} \, \frac{d R}{d t} \, \rho(R) . \label{eqn:bubble-vanish-rate}
	\end{align}
	Since $d N / d t < 0$ during coarsening, and $d R / d t$ takes a finite negative value at $R = 0$ for $\hat{\Pi} > 0$ (and $\phi > 0$) by eqn~\eqref{eqn:growth-law-bb}, we must have $\rho(0) > 0$ for our growth law. For Lemlich's law~\eqref{eqn:growth-law-lemlich}, $d R / d t$ diverges as $R \to 0$, and so $\rho(0) = 0$. The film area vanishes in this limit for $\phi > 0$ by eqn~\eqref{eqn:film-area-approx}, thus suppressing the singularity in our model due to the assumption of border blocking.
	
	Simulations\cite{2018-khakalo} and experiments\cite{2023-galvani,2025-galvani} have found a large population of small bubbles in coarsening wet foams, in 2D and 3D respectively. However, the forms of the observed bubble-size distributions (which are similar to each other) differ from eqn~\eqref{eqn:probability-dist}: they have large peaks at small bubble radius, which decrease in height as $\phi$ increases,\cite{2017-khakalo-pre-n,2018-khakalo,2023-galvani,2025-galvani}$^\text{,}$\footnote[8]{The preprint\cite{2017-khakalo-pre-n} of \citet{2018-khakalo} contains further plots of the simulated scaling-state distributions in the border-blocking case we consider.} whereas the value of the distribution at $R = 0$ increases with $\phi$ in our prediction (see Fig.~\ref{fig:coarsen-rate}). The differing form of our distribution also means that our predicted polydispersities $\mathcal{P}$ (see Figs.~\ref{fig:dist-props} and~\ref{fig:poly-v-rad}) differ from those observed in experiments (see the supplementary material of \citet{2023-galvani}): the large peak and long tail in their distributions at moderate $\phi$ result in a larger $\mathcal{P}$ than we find, while the reduction in the size of the peak with increasing $\phi$ causes $\mathcal{P}$ to decrease,\cite{2023-galvani} whereas we find it to increase.
	
	The experiments\cite{2023-galvani} are not wholly comparable with our model, since inter-bubble adhesion is believed to have been present in the former.\cite{2023-pasquet-b} However, such adhesion should increase the shrinkage rate of small bubbles due to the increased contact between bubbles,\cite{2021-feng,2023-galvani} and thus act to reduce the population of small bubbles. Also, we recall that our model neglects gas flow through the bulk liquid, and so we can only compare directly with the simulations reported by \citet{2018-khakalo} which also omit this contribution.\cite{2017-khakalo-pre-n}
	
	Both \citet{2018-khakalo} and \citet{2023-galvani} attribute the peaks in their distributions to the presence of rattlers, i.e.\ bubbles that are not compressed by their neighbours (though they may be adhered to them in the presence of a contact angle\cite{2023-galvani}). In the absence of adhesion, these bubbles lose contact with their neighbours, their films thus shrinking to zero size before their radius has done so (resulting in a border-blocking growth rate of zero).\cite{2000-gardiner,2018-khakalo} Our mean-field model does not predict rattlers of finite size, since $A / S > 0$ for all $R > 0$ by eqn~\eqref{eqn:film-area-approx}. However, rattlers might be present in the finite-element simulation results\cite{2024-morgan,2025-morgan-thesis} shown in Figs.~\ref{fig:growth-law-2d}(b) and~\ref{fig:growth-law-3d}(b), as small bubbles with (almost) zero growth rate.
	
	We now consider the possible roles of rattlers in a mean-field coarsening model. One possibility, suggested in 2D by the binned mean of Fig.~\ref{fig:growth-law-2d}(b), is that very small bubbles have nonzero growth rates, described by eqn~\eqref{eqn:growth-law-bb} or another growth law, when the scatter at fixed $R$ is averaged. In other words, there are no rattlers of finite size in an averaged description, and all rattlers that exist in a real foam are a result of scatter away from the average (such scatter is apparent in Figs.~\ref{fig:growth-law-2d} and~\ref{fig:growth-law-3d}) --- differing bubble environments mean that some small bubbles become rattlers, and cease to engage in coarsening (at least for a time\cite{2018-khakalo}), leading to an accumulation of rattlers and hence peaked distributions similar to those observed by \citet{2018-khakalo} and \citeauthor{2023-galvani}\cite{2023-galvani} It seems unlikely that such distributions could be captured by any mean-field model in this case.
	
	Another possibility is that the averaged growth rate does reach zero at a finite bubble size; i.e.\ rattlers exist in the mean field. Fig.~\ref{fig:growth-law-3d} suggests this in 3D, although possibly as an artefact of our aggregation of data from several small foams (see Section~\ref{sec:growth-law-bb}). A stronger argument comes from inspecting Fig.~\ref{fig:growth-law-2d}(a) closely: extrapolating the trend of the smallest cluster of bubbles (those with three neighbours\cite{2013-roth}) suggests that the growth rate reaches zero for $R > 0$ in 2D (indeed, this is predicted by the theory of \citet{2013-roth} and \citet{2017-schimming}). \citet{2023-galvani} argue for such a cutoff radius in 3D (below which bubbles are rattlers) based on the size of a Plateau-border junction, and their experimental results support this. However, rattlers would cease to evolve in the absence of adhesion or bulk gas transfer, and so no bubbles would vanish as a result of coarsening. Thus, a conventional scaling state would be precluded for such a border-blocking model without bubble adhesion. This does not contradict the experimental results of \citeauthor{2023-galvani},\cite{2023-galvani} due to the likely presence of adhesion (and bulk gas transfer).\footnote[9]{We note that the observed peaked distributions\cite{2018-khakalo,2023-galvani} can be reproduced qualitatively with the simulations described in Section~\ref{sec:methods} (at least during the coarsening transient), if $\hat{R}$ in the numerator of eqn~\eqref{eqn:film-area-approx} is replaced by $\max\{0, \hat{R} - c\}$ (for a small constant $c$), and a small multiple of the LSW law~\eqref{eqn:growth-law-lsw} is added to the resulting growth law (to roughly approximate bulk gas transfer;\cite{2018-khakalo} c.f.\ also ref.~\citenum{1972-ardell}). However, these adjustments are clearly ad hoc, and physical arguments are needed to fix their parameters.}
	
	The second possibility, that the averaged bubble growth rate reaches zero at finite radius, appears more likely to us. However, the simulations of \citet{2018-khakalo} (particularly the results included in their preprint\cite{2017-khakalo-pre-n}) suggest that there is a conventional scaling state when adhesion and bulk gas transfer are omitted, and hence that there is no cutoff size below which all bubbles are rattlers. Recalling Fig.~\ref{fig:growth-law-2d}(a), it is conceivable that the growth-rate distribution shown varies with polydispersity --- perhaps, as rattlers accumulate, the cutoff size decreases, thus ensuring their eventual disappearance. It may be fruitful to explore the relation between coarsening wet foams and the highly bidisperse soft disk systems studied by \citeauthor{2025-petit},\cite{2025-petit} who find a second jamming transition at $\phi < \phi_\text{c}$ for the small disks.
	
	It would be very useful for the development of mean-field models of coarsening to distinguish between these (and any other) possibilities, perhaps through further bubble-scale simulations. Three-dimensional finite-element simulations\cite{2021-wang,2025-morgan-thesis} of larger and more polydisperse foams (hence containing more small bubbles) could be performed, although improvement to our current numerics would be needed to make larger systems feasible for us. Two-dimensional finite-element simulations\cite{2014-kahara,2019-boromand,2024-morgan} of more polydisperse foams may also be useful, as might simulations tracking the evolution under coarsening of small bubbles. For comparability, all these simulations should be without bubble adhesion.
	
	Bubble-model\cite{1995-durian,1999-gardiner,2000-gardiner,2018-khakalo} simulations in 2D or 3D would seem to be a promising approach, due to the feasibility of simulating large wet foams. It would be interesting to see whether scaling-state bubble-size distributions with rattlers omitted are similar to our prediction~\eqref{eqn:probability-dist}. Further wet-foam coarsening experiments, such as those described by \citeauthor{2023-galvani},\cite{2023-galvani} in which individual bubbles can be tracked, may also be helpful in clarifying the dynamics of rattlers.

\section{Conclusion}\label{sec:conclusion}
	
	In the present work, we derived\cite{1959-lifshitz,1983-marqusee} the scaling-state bubble-size distribution~\eqref{eqn:probability-dist} predicted by our proposed mean-field border-blocking growth law~\eqref{eqn:growth-law-bb} (which we validated against averaged data from bubble-scale simulations\cite{2024-morgan,2025-morgan-thesis}), for foams with zero contact angle and at any liquid fraction $\phi < \phi_\text{c}$. We showed that the scaling-state growth exponent is $\alpha = 1/2$ at all $\phi < \phi_\text{c}$, as expected since the model incorporates only gas transfer through contact films between bubbles.\cite{2023-pasquet-b} Using mean-field simulations (with methods adapted from \citet{1997-de-smet}) in which bubbles evolve according to the proposed growth law~\eqref{eqn:growth-law-bb}, we also checked that the derived scaling state is approached from various initial conditions (we discuss some apparent exceptions in the \hyperref[sec:app]{Appendix}, which are judged to be artefacts of the mean-field approach\cite{1989-brown,2022-li,2022-svoboda}).
	
	We then compared our predicted bubble-size distribution~\eqref{eqn:probability-dist} with prior experimental results\cite{2023-galvani,2025-galvani} and simulations using the bubble model.\cite{2018-khakalo} While, consistently with these, eqn~\eqref{eqn:probability-dist} reproduces a large population of small bubbles for $\phi > 0$, it exhibits qualitative differences from the previous observations, in which there is a large peak in the distribution at small bubble radii.\cite{2018-khakalo,2023-galvani,2025-galvani} This peak has previously been attributed to rattlers\cite{2018-khakalo,2023-galvani} --- small bubbles not pressed into contact with their neighbours --- whereas our model does not incorporate rattlers of finite size. The addition of rattlers would be an important refinement, which could be developed using the results of future bubble-scale simulations as discussed in Section~\ref{sec:discussion}. Nevertheless, we suggest that our predicted distribution~\eqref{eqn:probability-dist} serves as an initial approximation incorporating variation with $\phi$ (noting that any refinements to our model would likely preclude an analytical derivation), which should allow the qualitative influence of additional effects such as gas transfer through bulk liquid\cite{2012-fortuna,2017-schimming,2023-durian-pre-n-yo} or finite contact angle\cite{2018-cox,2023-pasquet-b} to be studied by adding approximations for these to the growth law~\eqref{eqn:growth-law-bb}. The numerical methods\cite{1997-de-smet} of Section~\ref{sec:methods} can be straightforwardly adapted to use other growth laws. A model for bulk gas transfer may also allow estimation of the liquid fraction above which the border-blocking assumption is a poor approximation, though this will vary with\cite{2017-schimming,2023-pasquet-b,2023-durian-pre-n-yo} $h / \langle R \rangle$. Furthermore, it may be fruitful to study whether the tail of the bubble-size distribution is sensitive to the form of the growth law at small radii.\cite{1989-brown}
	
	A version of eqn~\eqref{eqn:growth-law-bb} augmented with an approximation for bulk-liquid gas transfer\cite{2017-schimming,2023-durian-pre-n-yo} could be used to predict the variation of growth exponent $\alpha$ with $\phi$, by adapting the approach of \citeauthor{1979-white}.\cite{1979-white} This may provide a prediction without fitting parameters, thus developing the existing model of \citeauthor{2023-durian-pre-n-yo}.\cite{2023-durian-pre-n-yo}

\section*{Data availability}

	The data and simulation code described in this work (including the 3D finite-element simulations\cite{2025-morgan-thesis}) are available from the Aberystwyth University Research Portal (\url{https://doi.org/10.20391/59c8e4a2-08ce-40ce-b3e7-0b7befdcd3f4}).
	
\appendix

\section*{Appendix}\label{sec:app}
	
	Here we discuss convergence of our mean-field simulations (see Section~\ref{sec:sims}) with respect to time-step size $\Delta \tau$, along with additional apparent scaling states which arise from pathological initial conditions.
	
\subsection*{Simulation convergence}\label{sub:app:con}
	
\begin{figure}[h]
		\centering
		\includegraphics[width=\columnwidth]{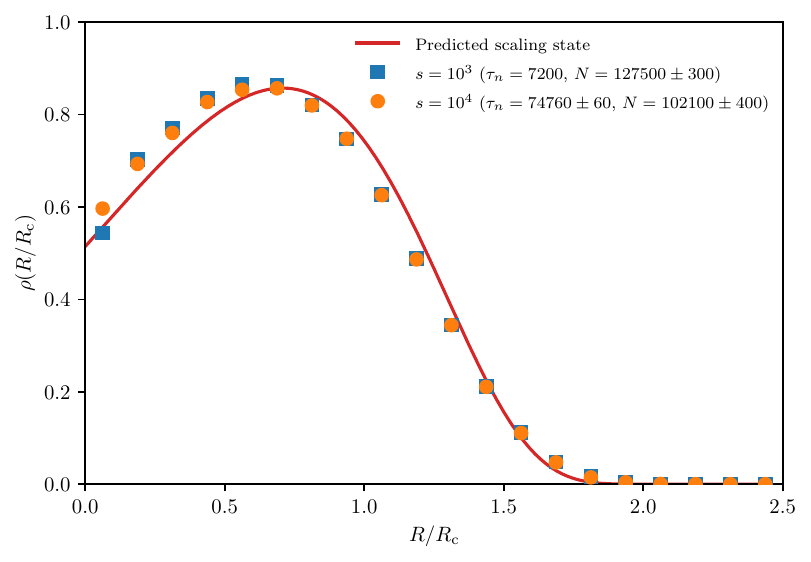}
		\caption{Bubble-size histogram when approximately $10^5$ bubbles remain in simulations of the type described in Section~\ref{sec:sims}. The liquid fraction is $35\%$, and results are given for two values of the parameter $m$ defining the time step $\Delta \tau$ (see Section~\ref{sec:methods}). Larger $m$ corresponds to smaller $\Delta \tau$. As in Fig.~\ref{fig:sim-dist-v-liq-frac}, the two histograms give the average over five distinct runs (the sample standard deviations are smaller than the markers). The initial bubble-size distribution was lognormal with $\sigma_R = 1/10$ in both cases. The predicted scaling-state distribution~\eqref{eqn:probability-dist} is shown for comparison.}
		\label{fig:converge}
	\end{figure}

	We recall the definition of $\Delta \tau$ from Section~\ref{sec:methods}: $\Delta \tau$ is such that $m$ steps of this size would be needed for a hypothetical bubble initially with the mean volume to vanish if its shrinkage rate were fixed at the current root-mean-square growth rate. We set $m = 10^3$ in Section~\ref{sec:methods}. Fig.~\ref{fig:converge} shows the effect on the bubble-size histograms of increasing $m$ by a factor of $10$ (i.e.\ decreasing $\Delta \tau$ by the same factor), from which we see little change in all bins except that for smallest radius $R$ (the plot is for liquid fraction $\phi = 35\%$ since large $\phi$ corresponds to the largest population of small bubbles by Fig.~\ref{fig:coarsen-rate}). To interpret this latter variation, we recall from Section~\ref{sec:methods} that $\Delta \tau$ determines the size below which bubbles are deleted (and their gas redistributed): the minimum volume $v$ is that for which the bubble would vanish at its present shrinkage rate after a time of exactly $2 \Delta \tau$. Hence, the size threshold increases with $\Delta \tau$ --- if the time step is larger, bubbles will be deleted earlier in their shrinkage process, and so fewer bubbles will be present in the bin of smallest radius, in agreement with Fig.~\ref{fig:converge}. This also lowers the polydispersity of the distribution slightly (not shown). Excepting this minor effect on the smallest bin of the histogram, we judge our simulations to be converged with respect to $\Delta \tau$. Another method for deleting small bubbles (by waiting until their volume drops below zero, for example) might reduce the effect on the smallest bin.
	
\subsection*{Additional apparent scaling states}\label{sub:app:nonunique}
	
\begin{figure}[h]
		\centering
		\includegraphics[width=\columnwidth]{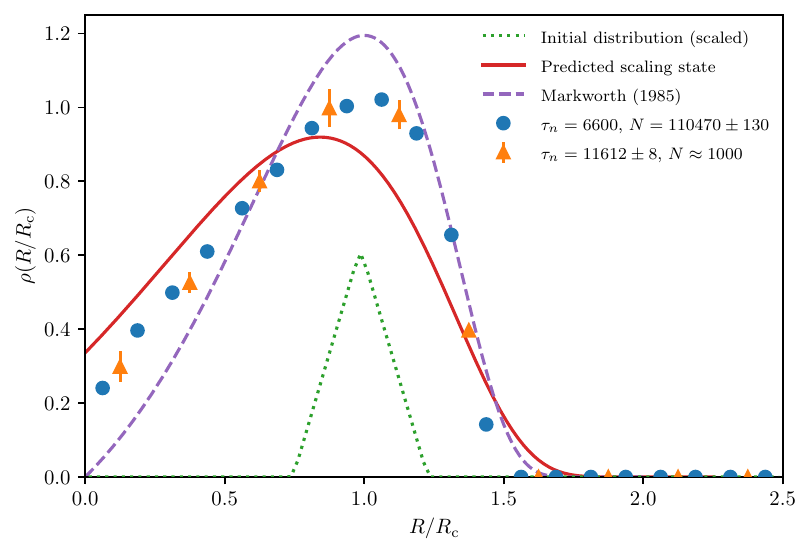}
		\caption{Bubble-size histograms for simulations with $\phi = 10\%$ and an initially triangular radius distribution. The latter is symmetric and has $\sigma_R = 1/10$. The data is presented in the same format as for Fig.~\ref{fig:sim-dist-v-liq-frac}.}
		\label{fig:alt-scaling-state}
	\end{figure}

	We now discuss some apparent exceptions we observe to the universality of the scaling state derived in Section~\ref{sub:reb:analysis}, which are interpreted to result from pathological initial conditions.\cite{1989-brown,2022-li,2022-svoboda} For a triangular initial bubble-size distribution, which is not continuously differentiable, we see from figure~\ref{fig:alt-scaling-state} that the system appears to approach a different scaling-state distribution. Like \citeauthor{1989-brown},\cite{1989-brown} who studied the LSW law~\eqref{eqn:growth-law-lsw}, we also find, in our simulations, that a discontinuous initial bubble size distribution (we used a uniform distribution) gives an apparent scaling state which likewise has a discontinuity (not shown). \citet{1989-brown} argued that the scaling state of the LSW law is not unique, and derived a family of scaling-state distributions. It is now believed\cite{2022-li,2022-svoboda} that, with the exception of the previously known scaling state of the LSW law,\cite{1959-lifshitz} the family found by \citet{1989-brown} consists of unstable scaling states. Indeed, when $\tau u$ is plotted against $\hat{R}$ (see Section~\ref{sub:reb:analysis}; plots not shown) for the apparent scaling state of Fig.~\ref{fig:alt-scaling-state}, two zeros are observed, with the apparent cutoff $\hat{R}_0$ near the smaller of these. Thus, this scaling state should be unstable to perturbation by the arguments of Section~\ref{sub:reb:analysis}.\cite{1983-marqusee,2022-svoboda} It appears that mean-field simulations of the type we use\cite{1989-brown,1997-de-smet} do not supply sufficient perturbations to the distribution during coarsening for the system to exit the spurious scaling state, at least over the timescale of our simulations.\cite{2022-li,2022-svoboda} Mean-field laws specify an equal growth rate for all bubbles of equal $R$, whereas real foams exhibit a large amount of scatter in the growth rates\cite{2007-lambert,2016-dittmann} (see our Figs.~\ref{fig:growth-law-2d} and~\ref{fig:growth-law-3d} for simulations thereof).
	
	We therefore neglect these apparent exceptions to the universality of the scaling state as artefacts of the mean-field growth law~\eqref{eqn:growth-law-bb} which would not arise in a real foam. Modelling approaches which add a diffusion term to the continuity equation~\eqref{eqn:continuity-eqn-n} may lack these artefacts.\cite{1977-binder,2019-zimnyakov}
	
\section*{Conflicts of interest}

There are no conflicts of interest to declare.

\section*{Acknowledgements}
	
	The authors acknowledge helpful discussions with Sylvie Cohen-Addad, Tudur Davies, Fran\c{c}ois Graner, Reinhard H\"ohler, and Rory McCranor. JM funded by Cronfa Ymchwil Joy Welch Research Fund and ESA REFOAM MAP contract 4000129502. SC partially supported by the Horizon 2020 Framework Programme for Research and Innovation, Grant Agreement number 101008140 EffectFact. JM and SC funded by EPSRC award number UKRI2390.


\balance

\renewcommand\refname{References}


\providecommand*{\mcitethebibliography}{\thebibliography}
\csname @ifundefined\endcsname{endmcitethebibliography}
{\let\endmcitethebibliography\endthebibliography}{}

\end{document}